\documentclass[5p,times]{elsarticle}
\usepackage{tabularx}  
\usepackage{cellspace}
\usepackage{listings}
\usepackage{fontawesome}
\usepackage{tcolorbox} 
\usepackage{fancyvrb}  
\usepackage{enumitem}  
\usepackage{graphicx}  
\usepackage{booktabs}  
\usepackage{xcolor}    
\usepackage{amssymb} 
\usepackage{pifont} 
\usepackage{fancybox} 
\usepackage{lipsum} 
\usepackage{bbding}
\biboptions{sort&compress}
\usepackage{mdframed}
\mdfdefinestyle{mystyle}{
  linewidth=1pt,
  linecolor=black,
  backgroundcolor=white,
}
\usepackage{wrapfig} 
\usepackage[section]{placeins}
\usepackage{booktabs}
\usepackage{amsmath,amssymb,amsfonts}
\usepackage[ruled,linesnumbered]{algorithm2e}
\usepackage{setspace}
\usepackage{graphicx}
\usepackage{comment}
\usepackage{float}
\usepackage{caption}
\usepackage{subcaption}
\usepackage{textcomp}
\usepackage{xcolor}
\usepackage{tcolorbox}
\usepackage{lipsum}
\usepackage{amsmath, bm}
\usepackage{listings}
\usepackage{amsmath, amssymb}
\usepackage{makecell}
\usepackage{multirow}
\usepackage{fancyhdr}
\usepackage{algorithmic}
\usepackage{adjustbox}
\usepackage[figuresright]{rotating}
\usepackage{colortbl}

\definecolor{highlight}{RGB}{196,216,242}

\definecolor{texgray}{RGB}{244, 245, 246}
\definecolor{texrightgray}{RGB}{99, 99, 99}
\definecolor{green}{RGB}{214, 230, 189}
\definecolor{leftgreen}{RGB}{93, 116, 162}
\definecolor{gray}{RGB}{200,200,200}
\definecolor{pinkback}{RGB}{242, 232, 227}
\newmdenv[
  leftline=false,
  rightline=true,
  topline=false,
  bottomline=false,
  backgroundcolor=texgray, 
  linecolor=texrightgray, 
  linewidth=2pt,
  skipabove=5pt
]{myrightline}

\newmdenv[
  leftline=true,
  rightline=false,
  topline=false,
  bottomline=false,
  backgroundcolor=pinkback, 
  linecolor=leftgreen, 
  linewidth=2pt,
  skipabove=5pt
]{myleftline}

\lstnewenvironment{code}[1][]
{\lstset{
  language=C,
  escapeinside={(*@}{@*)},
  basicstyle=\scriptsize\ttfamily,
  breaklines=true,
  numbers=none,
  numbersep=5pt,
  xleftmargin=10pt,
  showstringspaces=false, 
}}
{}

\usepackage{ifthen}
\newboolean{showcomments}
\setboolean{showcomments}{true}
\ifthenelse{\boolean{showcomments}}
  {\newcommand{\mynote}[2]{
    \fbox{\bfseries\sffamily\scriptsize#1}
    {\small$\blacktriangleright$\textsf{\emph{#2}}$\blacktriangleleft$}
   }
  }
  {\newcommand{\mynote}[2]{}
  }

\newcommand{\revise}[1]{\textcolor{black}{#1}}

\journal{Journal of \LaTeX\ Templates}

\begin{document}
\begin{sloppypar}
\newtheorem{definition}{Definition} 
\begin{frontmatter}

\title{Exploring the Potential and Limitations of Large Language Models for Novice Program Fault Localization
}

    \author[BUCT CIST]{Hexiang Xu}
	\author[BUCT CIST]{Hengyuan Liu$^*$}
	\ead{lhywandm@163.com}

	\author[BIPT]{Yonghao Wu}
	\author[BUCT CIST]{Xiaolan Kang}

	\author[NTU]{Xiang Chen}

	\address[BUCT CIST]{College of Information Science and Technology, Beijing University of Chemical Technology, Beijing, China}
	\address[BIPT]{College of Information Engineering, Beijing Institute of Petrochemical Technology, Beijing, China}
	\address[NTU]{School of Artificial Intelligence and Computer Science, Nantong University, Nantong, China}
	\author[BUCT CIST]{Yong Liu$^*$}
	\ead{lyong@mail.buct.edu.cn}

\begin{abstract}

Novice programmers often face challenges in fault localization due to their limited experience and understanding of programming syntax and logic. Traditional methods like Spectrum-Based Fault Localization (SBFL) and Mutation-Based Fault Localization (MBFL) help identify faults but often lack the ability to understand code context, making them less effective for beginners. In recent years, Large Language Models (LLMs) have shown promise in overcoming these limitations by utilizing their ability to understand program syntax and semantics. LLM-based fault localization provides more accurate and context-aware results than traditional techniques. \revise{This study evaluates six closed-source and seven open-source LLMs using the Codeflaws,} Condefects, and BugT datasets, with BugT being a newly constructed dataset specifically designed to mitigate data leakage concerns. Advanced models with reasoning capabilities, such as \revise{OpenAI o3 and DeepSeekR1}, achieve superior accuracy with minimal reliance on prompt engineering. In contrast, models without reasoning capabilities, like GPT-4, require carefully designed prompts to maintain performance. While LLMs perform well in simple fault localization, their accuracy decreases as problem difficulty increases, though top models maintain robust performance in the BugT dataset. Over-reasoning is another challenge, where some models generate excessive explanations that hinder fault localization clarity. Additionally, the computational cost of deploying LLMs remains a significant barrier for real-time debugging. 
\revise{LLM's explanations demonstrate significant value for novice programmer assistance, with one-year experience participants consistently rating them highly.}
Our findings demonstrate the potential of LLMs to improve debugging efficiency while stressing the need for further refinement in their reasoning and computational efficiency for practical adoption.

\end{abstract}

\begin{keyword}
Large Language Model, Fault Localization, Empirical Study, Novice Programming, Prompt Engineering
\end{keyword}
\end{frontmatter}

\section{Introduction}
\label{sec: Introduction}

In the process of learning programming, novice programmers often face significant challenges in fault localization. Due to their lack of experience with programming syntax, logic, and debugging techniques, novices struggle to identify and fix faults in their code. Programming faults can manifest at multiple levels, including syntax faults, logical faults, and runtime faults~\cite{li2021applying}. Each type of fault requires a deep understanding of the code to localize and correct effectively. Unlike natural languages, programming languages have stricter syntax rules and more rigid logical structures~\cite{tunstall2022natural}, which further complicates the learning curve and fault localization for beginners. This often results in novices spending excessive time on debugging, severely impacting their development efficiency.
Traditionally, fault localization relies on methods such as Spectrum-Based Fault Localization (SBFL)~\cite{li2021vsbfl} and Mutation-Based Fault Localization (MBFL)~\cite{9759520}. While these techniques can identify simple faults to some extent, they often lack a deep understanding of code context, particularly when dealing with complex faults. For novice programmers, the limitations of these tools are especially distinct: they provide insufficient intuitive guidance, failing to meet the needs of novices who require clear and comprehensible debugging support during their learning phase. Consequently, existing fault localization methods remain inadequate in helping novice programmers quickly and accurately localize and fix faults, highlighting the urgent need for more advanced tools to bridge this gap.

In recent years, the rapid development of artificial intelligence and deep learning technologies, particularly the emergence of Large Language Models (LLMs), has provided new possibilities for fault localization. Models such as GPT~\cite{DBLP:journals/rfc/rfc9405}, GLM~\cite{du2022glm}, and Llama~\cite{touvron2023llama} have demonstrated remarkable performance in natural language processing (NLP) tasks, including text generation, machine translation, and question-answering systems~\cite{JIM2024100059}. These models, pre-trained on vast amounts of data, possess the potential to understand and generate code, making them powerful tools for assisting in programming tasks. LLM-based fault localization methods leverage the models' understanding of program syntax and semantics to automatically localize faults in code, offering more precise and contextually relevant suggestions compared to traditional methods~\cite{yang2024large}. Moreover, these models can extend to code repair, not only helping programmers identify faults but also providing effective repair suggestions, thereby significantly enhancing debugging efficiency~\cite{sun2018search}.

Despite the promising advancements of LLMs in fault localization, several limitations hinder their practical application. One major issue is over-reasoning~\cite{chiang-lee-2024-reasoning}, where models generate excessive or unnecessary explanations that may obscure the actual fault. This not only increases the cognitive load for novice programmers but also introduces misleading information, making debugging less efficient. Additionally, the computational cost of deploying LLMs remains a significant challenge. Running LLMs requires substantial hardware resources, making them less feasible for real-time or resource-constrained environments. Lastly, the effectiveness of LLM's explanations for novice programmers remains to be investigated~\cite{DBLP:conf/emnlp/WangMC24}. Thus, despite the enormous potential of LLMs, their accuracy and reliability in fault localization for novice programs remain to be further validated~\cite{kang2023preliminary}.

To investigate these issues, we aim to empirically assess the performance of different LLMs across various datasets, evaluating their accuracy, efficiency, and usability in fault localization for novice programs. To ensure robust evaluation, we have constructed a new internal dataset to avoid data leakage~\cite{DBLP:journals/corr/abs-2410-06704}, and combined it with public datasets for parallel testing~\cite{wu2023condefects}.

This study evaluates the fault localization performance of six closed-source LLMs (e.g., OpenAI o3, GPT-3.5-Turbo, etc.) and seven open-source LLMs (e.g., DeepSeekR1, Llama3, etc.)  across Codeflaws, Condefects, and BugT datasets. All closed-source LLMs outperform traditional SBFL and MBFL methods, \revise{while DeepSeekV3 and DeepSeekR1 specialized in code tasks, surpasses most closed-source and open-source models. OpenAI o3 and DeepSeekR1 consistently achieve superior results compared to smaller models and traditional techniques. For models with strong modular reasoning, such as OpenAI o3 and DeepSeekR1, prompt engineering has minimal impact. In contrast, models like GPT-4 rely heavily on specific prompts, particularly novice program prompts, for effective fault localization. LLM accuracy decreases as problem difficulty increases in Codeflaws and Condefects, but top models maintain high accuracy even at peak difficulty in BugT, suggesting its lower complexity. }
\revise{Additionally, LLM generated fault explanations demonstrate significant value for novice programmer assistance, with one-year experience participants consistently providing high ratings across all dimensions, proving that LLMs offer important reference value in helping novice programs understand faults and conduct practical debugging.}

We significantly extend our previous conference paper~\cite{QRS2023LLM} with comprehensive advancements in multiple dimensions:

\begin{itemize}
\item \textbf{Extended Model Evaluation:} We incorporate \revise{eight} additional LLMs, including \revise{four} cutting-edge proprietary models (\revise{OpenAI o3}, o1-mini, o1-preview, GPT-4o) and \revise{four} widely-researched open-source models (ChatGLM4, \revise{DeepSeekR1} DeepSeekV3, Llama3) to examine different architectural approaches to fault localization.
\item \textbf{New Dataset Construction:} We introduce a new self-created dataset with real programming faults, aiming to mitigate data leakage concerns and establish a more reliable evaluation benchmark for assessing LLMs' capabilities.
\item \textbf{Difficulty Impact Analysis:} We examine how problem difficulty levels across three datasets affect LLMs' fault localization performance, providing insights into model behavior at varying complexity levels.
\item \textbf{Model Size and Performance Analysis:} We investigate cases in the Codeflaws dataset where smaller-scale models unexpectedly outperform larger ones, questioning common assumptions about the relationship between model size and fault localization effectiveness.
\item \textbf{User Study and Manual Analysis:} We analyze LLMs' fault explanations through manual examination and user surveys, gathering feedback on their educational value and practical usefulness for novice programmers in debugging scenarios.
\end{itemize}

The contributions of this work are summarised as follows:

    \begin{itemize}
    \item
    \textbf{Comprehensive evaluation of LLMs for novice programming fault localization}

    We systematically assess various LLMs across different prompting strategies and problem difficulty levels, revealing their unique advantages when handling novice programs.
    Additionally, our supplementary authentic novice program dataset helps mitigate potential data leakage concerns in LLM applications within this domain.
    \item
    \textbf{Analysis of LLMs' limitations and complementary relationships with traditional fault localization methods}
    
    Our comparative analysis reveals that model scale does not always guarantee superior performance, with certain closed-source models like o1 occasionally underperforming compared to GPT-3.5-Turbo despite having more advanced architectures.
    Through overlapping line analysis, we demonstrate that different methods successfully identify unique fault types that others miss, confirming that integrating multiple approaches can lead to more comprehensive fault localization solutions.
    
    \item
    \textbf{Empirical guidance for enhancing programming education feedback}
    
    Through questionnaire validation of LLM-generated fault explanations, we establish clear directions for improving automated feedback systems in programming education.
    Our findings confirm that LLMs' interpretable outputs offer significant pedagogical value, providing key insights for future educational tool development.
\end{itemize}

To facilitate future study, we share our source code and experimental data in a GitHub repository~\footnote{{https://github.com/Xucranger/PLofLBFL}}.

\section{Background}
\label{sec:Background}

\subsection{Programming Education}

As programming education evolves, fault localization has become crucial for novice programmers who struggle with understanding programming structures and debugging. 
Automated fault localization tools have significantly advanced, helping novice programmers quickly identify and fix faults by analyzing fault messages, stack traces, or execution paths. These tools enhance learning efficiency by providing immediate feedback and reducing instructors' workload.
Moreover, interactive platforms and intelligent tutoring systems integrate fault localization with real-time feedback, allowing novice programmers to develop debugging skills through practice. By optimizing these tools and teaching strategies, we can enhance novice programmers' coding abilities, reduce frustration, and accelerate their growth.

\noindent \textbf{\revise{1)Novice Programmer and Novice Program}}

\revise{Novice Programmer~\cite{10.1145/3639474.3640059} refers to students who are in the early stages of learning programming. As beginners, novice programmers usually lack an in-depth understanding of programming languages and have limited ability to debug errors independently. Novice Program~\cite{201537} refers to code written by novice programmers, usually in the early stages of learning programming. These programs usually contain multiple errors, reflecting the novice programmer's lack of understanding of the syntax and semantics of the programming language.}

\noindent \textbf{2) Novice Programming Assistance}

Novice programming assistance~\cite{penman2021does} is a crucial educational approach aimed at helping students develop their fundamental coding skills and enhance problem-solving abilities. By utilizing structured programming exercises, real-time feedback, and timely guidance, this method addresses the initial frustration often experienced by beginners, allowing them to better understand and apply core programming concepts. The approach emphasizes active learning, encouraging students to actively engage in discovering solutions through continuous experimentation and practice.
The primary goal of novice programming education is to gradually build students' abilities, starting from foundational coding skills and progressing to more complex problem-solving techniques. With the increasing availability of interactive platforms and AI-driven tools, modern programming education provides learners with personalized feedback and real-world problem-solving opportunities, creating a more dynamic and effective learning environment. As technology evolves, these advancements contribute to an enriched learning experience, helping students navigate the challenges of software development more effectively.

\noindent \textbf{3) Fault Localization on Novice Programs}

Fault localization~\cite{li2021clacer, LI2023111822,7390282} presents a significant challenge in novice programming education. Traditional techniques, such as breakpoint debugging and print statement outputs, commonly used in industrial software development, may not be well-suited for the educational context. These methods assume that the user possesses a certain level of problem-solving ability and an in-depth understanding of program execution flows, which novice programmers typically lack. As a result, while traditional debugging methods can provide some assistance, their complexity often limits their effectiveness for beginners. This highlights the necessity for developing fault localization tools and methods that are better aligned with the needs of novice learners, offering simpler and more intuitive solutions within the educational context.

\subsection{Fault Localization}

In software development, fault localization is a critical process for ensuring the proper functionality of the software, particularly in high-stakes domains where severe faults can lead to significant financial losses or even injuries~\cite{lou2020can}. Traditionally, fault localization has relied heavily on manual debugging, a time-consuming and resource-intensive task~\cite{jones2002visualization}. To address the inefficiencies of this approach, researchers have developed automated techniques such as Spectrum-based Fault Localization (SBFL) and Mutation-based Fault Localization (MBFL)~\cite{9759520,li2019empirical,wu2023identifying,liu2019weighted,arrieta2018spectrum,papadakis2015metallaxis,wang2020ietcr,Liu2024Delta4MsIM,li2020hmer,liu2018optimal}. These methods enhance the fault localization process by capturing dynamic execution data, scoring code statements based on their likelihood of being faulty, and generating ranked lists of potential fault locations. This not only reduces the time required for debugging but also improves the accuracy of identifying faults.

\noindent \textbf{1) Spectrum-based Fault Location}

SBFL (Spectrum-Based Fault Localization)~\cite{arrieta2018spectrum} is an automated method for identifying faulty code statements by analyzing the results of test cases~\cite{keller2017critical}. It is based on the idea that statements executed by more failing tests are more likely to be faulty, meaning that both passed and failed test outcomes are crucial for the analysis~\cite{lou2020can}. SBFL tracks the execution paths of test cases, converts them into coverage data, and calculates a ``suspiciousness" value for each statement using various formulas, including Jaccard~\cite{jaccard1901etude}, Ochiai~\cite{abreu2006evaluation}, Dstar*~\cite{wong2013dstar}, Op2~\cite{naish2011model}, among others. These suspiciousness values are then used to rank the statements according to their likelihood of being faulty, thereby helping developers efficiently localize the source of faults.

\noindent \textbf{2) Mutation-Based Fault Localization}

Mutation-Based Fault Localization (MBFL) is a method for locating program faults through mutation testing. The core idea is to introduce small mutations (mutants) into the program and use existing test cases to test these mutants. If a mutant is killed by a failing test case, the corresponding code line is considered potentially faulty. The main steps of MBFL include: first, Leveraging the coverage data derived from failed test executions, systematically generating mutation variants targeting the statements exercised by the test cases; then, running these mutants and recording the test case results; and finally, calculating the suspiciousness of each code line based on the mutant outcomes. By ranking the suspiciousness of each line, MBFL helps developers efficiently localize faults. This method can improve fault localization accuracy, especially in complex systems, and shows great potential in practical applications.

\noindent \textbf{3) Large Language Model Fault Localization}

Fault localization using LLMs refers to the utilization of LLMs, such as GPT, Llama, and others, to assist in identifying and correcting various faults in programs, including syntax faults, logical faults, and runtime faults. By analyzing the context of the code, LLMs can offer suggestions for fault localization, helping developers quickly identify faulty lines, and even make inferences in more complex cases. In addition to fault localization, some advanced models can also generate repair suggestions~\cite{lohr2024yourenottype} and automate debugging processes, reducing the time developers spend on troubleshooting. These models are typically trained on extensive open-source code and programming data, allowing them to provide accurate localization and repair suggestions for common programming issues.

\subsection{Large Language Model}

Large Language Models (LLMs) are deep learning models characterized by their vast number of parameters, often in the billions, and trained on extensive datasets~\cite{zhang2023critical}. These models have demonstrated exceptional performance in understanding and generating natural language, making them indispensable in a wide range of language-related tasks such as translation, content generation, and summarization~\cite{tunstall2022natural}.
In recent years, LLMs have also made significant strides in the domain of programming \revise{languages~\cite{10.1145/3715007}. }Their application in software engineering is rapidly expanding, with LLMs excelling in tasks like code migration, code translation, and code completion. In some scenarios, these models outperform traditional methods, showcasing their immense potential to revolutionize software development processes~\cite{tian2023ChatGPT, sobania2023analysis, sridhara2023ChatGPT, xie2023chatunitest}.

\noindent \textbf{1) Data leakage in LLMs}

\begin{figure}[htp]
    \centering
    \includegraphics[width=0.48\textwidth]{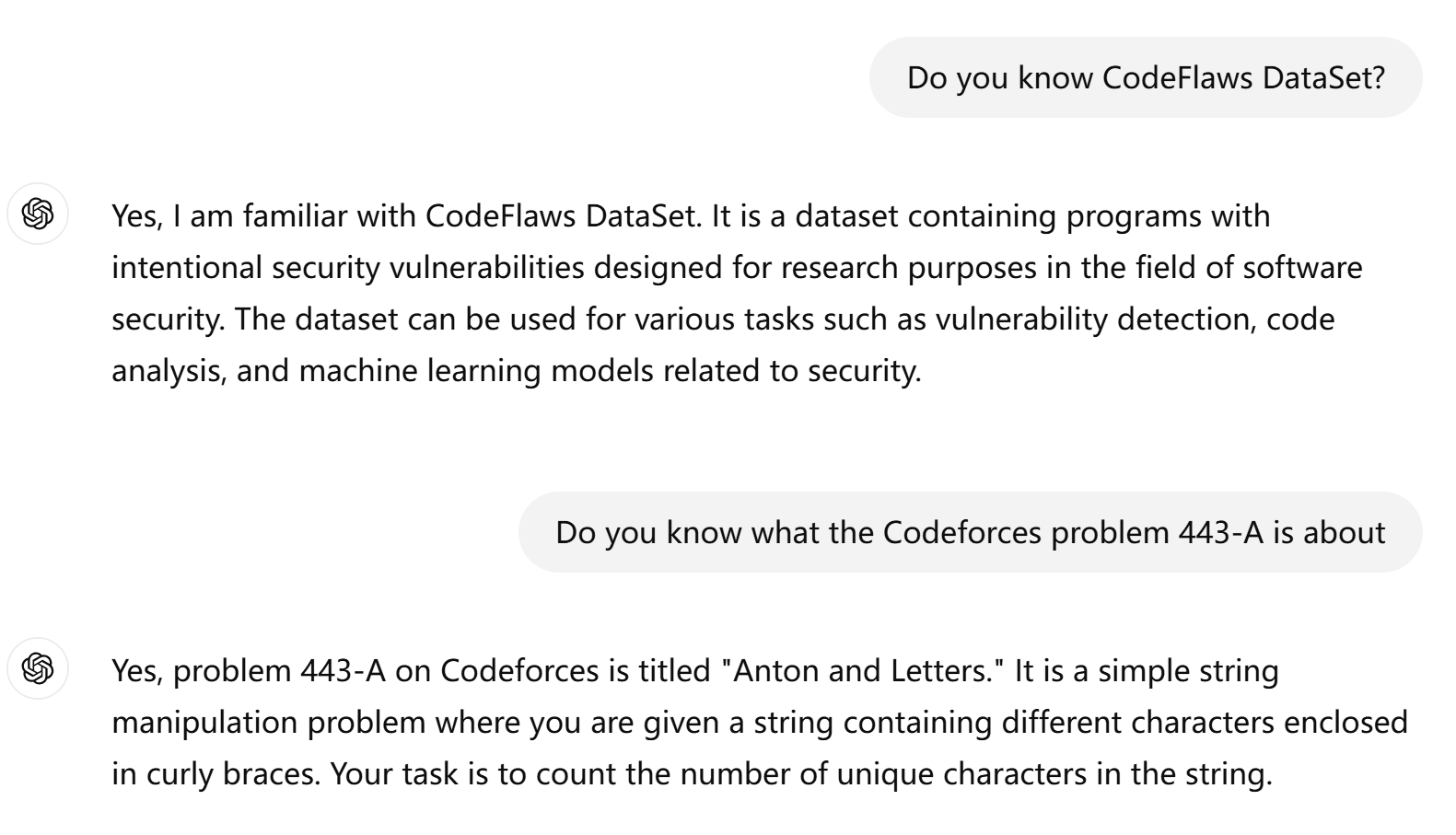}
    \caption{GPT-3.5-Turbo's perception of the Codeflaws dataset.}
    \label{fig:A conversation with GPT-3.5-Turbo}
\end{figure}

Data leakage~\cite{zhou2025lessleakbenchinvestigationdataleakage} in LLMs refers to a situation where the model inadvertently accesses test data or target data during the training process, leading to overly optimistic performance results that impact the model’s ability to generalize in real-world applications. Data leakage can arise from implicit future information in the training data or improper data partitioning during processing. It not only causes inaccurate evaluation results but can also pose privacy and security risks. As LLMs are applied across various domains, preventing data leakage has become a critical issue for enhancing model reliability and security.
Figure~\ref{fig:A conversation with GPT-3.5-Turbo} demonstrates how the offline version of GPT-3.5-Turbo perceives the Codeflaws dataset through a simulated UI conversation.

\noindent \textbf{2) Reasoning Capabilities of Large Language Models}

\revise{Reasoning in LLMs refers to the computational ability to decompose complex problems into a sequence of logically coherent intermediate steps to arrive at valid solutions~\cite{wei2022chain}. The effectiveness of reasoning is positively correlated with model scale and is significantly enhanced when models are provided with explicit step-by-step demonstrations through Chain-of-Thought (CoT) prompting techniques. Mainstream LLMs implement reasoning through different paradigms: Transformer-based models like GPT series~\cite{openai2024gpt4osystemcard} use autoregressive generation with attention mechanisms, while recent models like OpenAI's o1~\cite{openai2024o1systemcard} series introduce explicit ``thinking" phases for internal reasoning before response generation.}

\section{\revise{Experiment Design}}
\label{Experimental Design}

\subsection{Dataset}


We select the Codeflaws~\cite{tan2017codeflaws}, Condefects~\cite{wu2023condefects}, and BugT datasets, all of which are sourced from online programming platforms. These datasets cover a wide range of logical faults, from incorrect conditional statements to array or list out-of-bounds access. They have been commonly used in education to teach debugging and fault localization techniques, and are also widely used in academic research to evaluate novel defect localization technologies~\cite{DBLP:conf/sigsoft/WuLZ024}.


\revise{The Codeflaws, BugT and Condefects datasets are publicly available open-source datasets. All programs used in our study come from educational contexts and is either publicly available or submitted for academic purposes. Therefore, no commercial or proprietary code was involved, and there are no privacy or intellectual property risks in our setting.}

\begin{itemize} 
  \item 
  \textbf{Codeflaws} 
The Codeflaws dataset consists of a curated collection of 3,902 unique C programs, derived directly from the Codeforces dataset. This includes both single-fault and multi-fault programs. On average, each program in the Codeflaws dataset contains about 36 lines of code and is associated with 43 test cases.
  \item 
  \textbf{Condefects} 
The Condefects dataset includes 1,254 Java programs and 1,625 Python programs, all containing flaws. These programs are sourced from the AtCoder online competition platform and are generated in September 2023. The dataset is specifically designed to reduce overlap with existing training data used for LLMs.
  \item
\textbf{BugT} 
The BugT dataset contains 7,097 C programs and 22,547 C++ programs and 10,507 Python programs, all sourced from the BuctOJ online programming platform, which serves as an educational programming platform for students at the Beijing University of Chemical Technology and is deployed on an internal network, completely eliminating the possibility of data leakage to LLMs.

\revise{The BugT dataset is constructed following the quality control procedures established by the Codeflaws dataset. During the data filtering process, approximately 6,594 redundant programs are removed. Specifically, on the BuctOJ online programming platform, users often submit multiple solutions for the same problem. From these submissions, we select the compilable correct version and compare it with the user's other submissions. Programs that differed by only a single line and are also compilable are retained, ensuring the construction of high-quality samples for fault localization research.}
\end{itemize}

\begin{table}[htbp]
  \centering
  \caption{Summary of Codeflaws and Condefects, BugT Datasets}
    \begin{tabular}{llccc}
    \toprule
    \multicolumn{2}{c}{\textbf{Datasets}} & \textbf{BugT} & \textbf{Codeflaws} & \textbf{Condefects} \\
    \midrule
    \multicolumn{2}{l}{\textbf{Program Used}} & 503   & 503   & 503 \\
    \multicolumn{1}{c}{\multirow{3}[0]{*}{\begin{sideways}\textbf{\#LOC}\end{sideways}}} & \textbf{Min.} & 9     & 8     & 6 \\
          & \textbf{Avg.} & 33    & 36    & 33 \\
          & \textbf{Max.} & 314   & 144   & 164 \\
    \multicolumn{2}{l}{\textbf{Test Case}} & 34    & 43    & 36 \\
    \multicolumn{2}{l}{\textbf{Program Language}} & C++   & C     & Java \\
    \multicolumn{2}{l}{\textbf{Program Total}} & 22547 & 3902  & 1254 \\
    \bottomrule
    \end{tabular}%
  \label{tab:DatasetInfo}%
\end{table}%


\revise{The programs included in the three datasets are all collected from online programming platforms. These platforms primarily target novice programmers, featuring problems that focus on basic algorithms. Consequently, the submitted code generally does not involve advanced design patterns or industrial-level programming practices, thus exhibiting typical characteristics of novice programs. To ensure that the selected programs are representative of novice-level programs, we prioritize problems with lower difficulty levels. However, to explore the impact of problem difficulty on fault localization performance, we deliberately include a small number of moderately difficult problems. Overall, the three datasets selected in this study are well aligned with the definition of novice programs.}

All of these three datasets utilize system-level testing methods, with test cases provided in an input-output format. Detailed information about these datasets is presented in Table~\ref{tab:DatasetInfo}, which includes the number of programs used in each dataset, the minimum, average, and maximum lines of code (LOC) per program, the number of test cases associated with each dataset, the programming language used, and the total number of programs across all datasets. 

We exclude programs from Codeflaws that could not be compiled or are too large to process due to token limitations, a common issue in LLM-based fault localization studies~\cite{kang2023preliminary,qin2024agentfl}. This results in a final set of 503 programs in the Codeflaws dataset. To ensure fairness in the experimental comparisons between the Codeflaws and other datasets, we randomly select an equal number of programs from the Condefects dataset and BugT dataset for the experiments. Additionally, the dataset includes C language programs from Codeflaws, Java language programs from Condefects, and C++ language programs from BugT. This not only facilitates a comparison of effectiveness across different programming languages but also demonstrates the transferability of our approach across languages.



\subsection{Baselines}


For benchmark comparison, we choose two traditional fault localization methods: Spectrum-Based Fault Localization (SBFL)~\cite{keller2017critical} and Mutation-Based Fault Localization (MBFL)~\cite{li2020hmer}. SBFL calculates the suspiciousness of code elements by analyzing test case coverage during program execution, whereas MBFL enhances this by introducing code mutations to simulate fault impacts, which helps improve fault localization accuracy. Both methods have been extensively validated in traditional software engineering, making them suitable benchmarks for evaluating the performance of LLMs in locating faults within novice-level programs. 

\subsection{Evaluation Metric}

\textbf{Top-N}
 is a widely recognized metric for evaluating fault localization techniques, designed to quantify their accuracy~\cite{lou2021boosting, wu2020fatoc}. Specifically, it measures how many faulty statements are identified within the top N entries of the suspiciousness ranking. Studies suggest that developers typically review only the statements ranked within the top 5 positions of the list~\cite{kochhar2016practitioners}. Consequently, our research emphasizes the Top-1 to Top-5 range.

\subsection{\revise{Experiment Setup}}


Within this study, we analyze novice-level programs derived from the datasets Codeflaws, Condefects, and BugT. These defective programs are input into various LLMs, including \revise{OpenAI o3}, o1-preview, o1-mini, GPT-4o, GPT-3.5-Turbo, GPT-4, ChatGLM4, ChatGLM3, \revise{DeepSeekR1}, DeepSeekV3, Llama2, Llama3, and Code Llama, utilizing tailored prompts specific to our research objectives. The aim is to evaluate the fault localization capabilities of these models. \revise{The ground truth labels for fault localization are the line numbers of the buggy line in fault programs. In order to reduce the experimental error caused by the randomness of the LLM output, we repeated each experiment 5 times and then took the average results.}  The results produced are subsequently gathered, refined, and subjected to statistical analysis to derive meaningful insights.


In the rest of this subsection, we show the experimental configurations and provide an in-depth discussion of the prompt engineering process.

\noindent \textbf{1) Large Language Models Setup}

\begin{table*}[htbp]
  \centering
  \caption{Details of Large Language Models (LLMs) Evaluated in Our Study}
    \begin{tabular}{lccccccc}
    \toprule
    \multicolumn{1}{c}{\textbf{Technique}} & \textbf{Developer} & \textbf{Release Date} & \textbf{Type} & \textbf{Training Data (up to)} & \textbf{Size}  & \textbf{\revise{Reasoning Capabilities}} \\
    \midrule
    \textbf{\revise{OpenAI o3}} & \revise{OpenAI} & \revise{2025.04} &  \revise{Closed-Source} & \revise{2024.06} & -   & \revise{83.3\%} \\
    \textbf{o1-preview} & OpenAI & 2024.09 &  Closed-Source & 2023.10 & -   & \revise{75.7\%} \\
    \textbf{o1-mini} & OpenAI & 2024.09 &  Closed-Source & 2023.10 & -      & \revise{-} \\
    \textbf{GPT-4o} & OpenAI & 2024.05 &  Closed-Source & 2023.10 & -     & \revise{56.1\%} \\
    \textbf{GPT-4} & OpenAI & 2024.04 &  Closed-Source & 2023.03 & -      & \revise{-} \\
    \textbf{GPT-3.5-Turbo} & OpenAI & 2022.11 &  Closed-Source & 2022.11 & -      & \revise{-} \\
    \midrule
    \textbf{ChatGLM4} & ZhiPuAI & 2024.09 & Open-Source & 2022.- & 9B        & \revise{18.01\%} \\
    \textbf{ChatGLM3} & ZhiPuAI & 2023.01 & Open-Source & 2022.01 & 6B        & \revise{-} \\
    \textbf{\revise{DeepSeekR1}} & \revise{DeepSeek} & \revise{2025.01} & \revise{Open-Source} & \revise{2023.10} & \revise{671B}      & \revise{71.5\%} \\
    \textbf{\revise{DeepSeekV3}} & DeepSeek & 2024.11 & Open-Source & 2023.10 & 671B      & \revise{64.8\%} \\
    \textbf{Llama3} & Meta  & 2024.4 & Open-Source & 2023.03 & 7B         & \revise{50.5\%} \\
    \textbf{Llama2} & Meta  & 2023.7 & Open-Source & 2022.07 & 7B        & \revise{-} \\
    \textbf{Code Llama} & Meta  & 2023.8 & Open-Source & 2023.07 & 7B        & \revise{-} \\
    \bottomrule
    \end{tabular}%
  \label{tab:Introduction of the LLMs involved in our experiment}%
\end{table*}%

In this empirical study, we employ various LLMs and introduce their features accordingly in Table~\ref{tab:Introduction of the LLMs involved in our experiment}. \revise{ The Reasoning capabilities score comes from the GPQA Diamond score on the Vellum AI Leaderboard~\footnote{{https://www.vellum.ai/open-llm-leaderboard}}~\footnote{{https://www.vellum.ai/llm-leaderboard}}. GPQA Diamond~\cite{rein2023gpqagraduatelevelgoogleproofqa} is a rigorous graduate-level evaluation framework specifically designed to assess advanced scientific reasoning capabilities. It consists of challenging multiple-choice questions in biology, physics, and chemistry, crafted by domain experts with PhD-level expertise. This high difficulty and scientific rigor make GPQA Diamond a highly authoritative indicator for evaluating the complex reasoning abilities required for advanced scientific inquiry, justifying its use as a reliable metric for LLM reasoning performance in our study.}
    
    \textbf{ChatGPT:}
    ChatGPT~\cite{kalla2023study}, \revise{ developed by OpenAI represents one of the leading closed-source LLMs, widely recognized for its advanced capabilities in both language understanding and generation. Although the model itself is not open-source, OpenAI provides access through its official API, supporting a range of configurations. In this study, we use the default settings of the API. As of July 2025, the latest versions include GPT-3.5-Turbo, GPT-4, GPT-4o, o1-mini,  o1-preview, and OpenAI o3, which offer improvements in multimodal support and efficiency. Among these, the OpenAI o3 is seen as a step towards a more flexible and customizable model architecture. By enhancing the model's training process and adding task-specific fine-tuning, OpenAI o3 demonstrates greater customization capabilities when handling complex tasks. The training data cutoff for GPT-3.5-Turbo is approximately September 2021, aligning with the development trajectory of OpenAI’s previous versions. From GPT-3.5-Turbo to OpenAI o3, OpenAI has continually enhanced the model's understanding abilities, generation quality, and computational efficiency, while making significant strides in personalization, customization, and multimodal support.}

    \textbf{ChatGLM:}
    ChatGLM is an open-source LLM series developed by the Chinese Zhipu AI team, aimed at enhancing the capabilities of Chinese natural language processing. Similar to GPT, ChatGLM adopts a Transformer-based autoregressive architecture, focusing on Chinese text generation and understanding, and supports multimodal tasks. ChatGLM-3~\cite{du2022glm}, is its third-generation model, with strong capabilities in Chinese dialogue generation and understanding, suitable for various NLP tasks. ChatGLM-4 further improves in terms of scale, performance, and multi-task processing abilities, optimizing dialogue fluency and long-text handling, especially demonstrating stronger advantages in the Chinese language context. In this study, we use chatglm4-8B and chatglm3-6B, which demonstrate high-quality linguistic fluency and naturalness, capable of generating coherent and contextually relevant text.
    
    \textbf{DeepSeek:}
    \revise{DeepSeekV3 and DeepSeekR1, developed by the DeepSeek team, represent significant advances in large language model architecture. DeepSeekV3 features a Mixture-of-Experts (MoE) architecture that dynamically activates selected expert networks to enhance computational efficiency and inference performance while reducing training instability. DeepSeekR1, the latest iteration, incorporates advanced reasoning capabilities and enhanced chain-of-thought reasoning mechanisms. Both models are optimized for code understanding and generation, pre-trained on large-scale, high-quality code datasets, and incorporate structure-aware mechanisms for better code logic analysis. Leveraging their respective architectural innovations, both DeepSeekV3 and DeepSeekR1 efficiently allocate computational resources for tasks like fault localization, improving both accuracy and efficiency in code-related applications.}
    
    \textbf{Llama:}
    Llama 2~\cite{touvron2023llama}, created by Meta AI, is an open-source LLM designed for general-purpose language understanding and generation tasks across various domains. Its scalable architecture allows fine-tuning for specific applications, and it has been widely adopted due to its transparency and accessibility. In this study, the Llama2-7B model is utilized, which balances performance and computational efficiency. Llama 3 is the anticipated next iteration of the Llama series, developed by Meta AI, which aims to further enhance language understanding and generation capabilities. While the exact features and architecture of Llama 3 are yet to be fully disclosed, it is expected to build on the successes of Llama 2 by improving scalability, accuracy, and efficiency. Llama 3 is likely to incorporate advancements in multimodal processing, fine-tuning for specialized tasks, and increased computational performance to handle larger datasets and more complex applications. 

    \textbf{Code Llama:} 
     Code Llama~\cite{roziere2023code}, an extension of the Llama 2 series tailored for programming tasks, provides enhanced functionalities in code generation, understanding, completion, and debugging. Trained on extensive code-focused datasets, it is a robust open-source LLM for developers. This study utilizes the CodeLlama-7B-Instruct-hf model, which is optimized for instruction-based coding scenarios. As of 2025, Code Llama maintains its position as a powerful tool for both academic and industrial code-centric applications, with active updates enriching its ecosystem.

\noindent \textbf{2) Prompt Design}

We developed a specialized prompt for input into various LLMs to enable fault localization in defective novice programs.
The design of the prompt integrates several essential components, outlined below.

   \textbf{Novice Description:} Building on the insights of Liu~\cite{liu2023pre}, integrating a fictional character role into prompts has proven effective in improving LLMs' comprehension of targeted tasks. By assigning a persona with specific expertise, we can shape the model’s responses to better address our research questions. As such, the prompt includes a detailed description of the novice program’s context, instructing the LLM to assume the role of an algorithm teacher specializing in fault localization for novice programs. The proposed scenario is as follows: a student has developed a novice program that fails to pass all test cases and approaches the algorithm teacher—renowned for expertise in fault localization—for assistance in identifying the problematic parts of the program.

    \textbf{Intent:} This method draws inspiration from the findings of Gao et al.~\cite{gao2023self}, which demonstrated that prompting LLMs to self-explain their responses significantly improves their comprehension of the questions posed. In this study, the prompts are designed not only to elicit results from the LLM but also to require the model to articulate the intended purpose of the target code, thereby enhancing the interpretability of its responses.

    \textbf{Reason:} Within the prompt, in addition to providing fundamental fault localization instructions, we require the LLM to include a justification for each suspected fault identified in its returned list. This requirement goes beyond the functionality of conventional fault localization approaches. By demanding detailed explanations for each identified line, our approach seeks to significantly deepen the LLM’s understanding of fault characteristics, thereby enhancing the accuracy of fault localization in novice programs.

    \textbf{Chain of Thought (CoT):} 
    Chain-of-Thought (CoT), as introduced by Wei et al.~\cite{wei2022chain}, has been demonstrated to significantly improve LLMs' ability to interpret and address questions. This approach entails dividing the problem into multiple sequential steps, enabling the LLM to reason systematically toward the final solution. In the context of the prompt developed for this study, CoT is implemented by segmenting the novice program fault localization task into two specific phases: first, guiding the LLMs to identify all lines that may contain faults, and second, prompting the models to compare and evaluate these lines, ranking them based on their suspiciousness values in descending order.

    \textbf{Sort:} In the final stage of the prompt, we specifically direct the LLM to rank its outputs based on the level of suspiciousness, ordering the results from most to least suspicious. This instruction not only sharpens the model's grasp of the fault localization task in novice programs but also ensures that the outputs are systematically organized, thereby improving the overall accuracy and efficiency of the fault localization process.

    During the experiment, we present the LLMs with the prompt, which includes the complete source code of a novice program, a concise yet thorough set of instructions for the LLMs, and integrates all the essential components outlined earlier.

\begin{figure}[htpb]
  
    \begin{subfigure}[b]{0.48\textwidth}
        \centering
        \includegraphics[width=\textwidth]{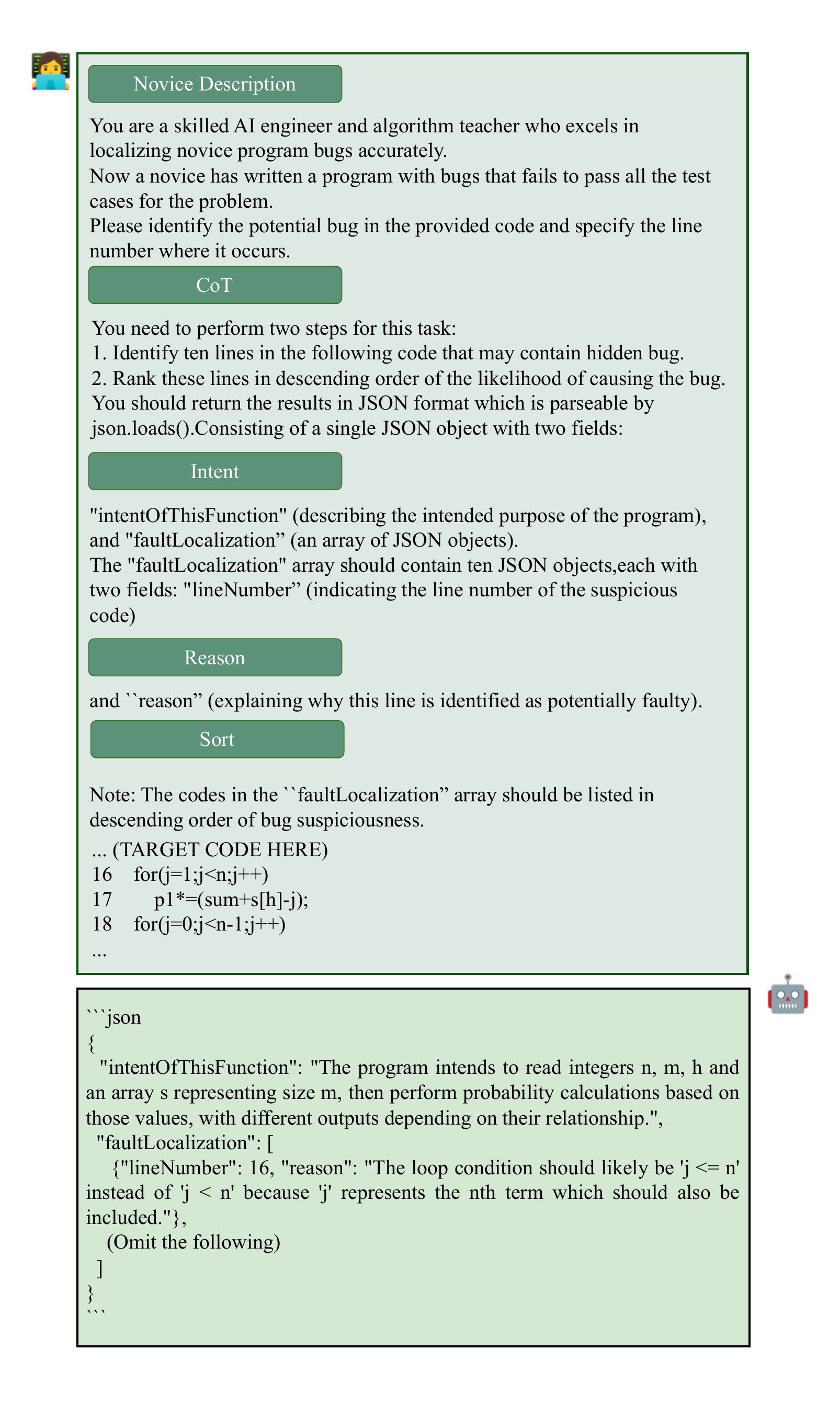}

    \end{subfigure}
    \caption{Example of the Prompt and the Expected Response}
     \label{fig:exampleofprompt}
\end{figure}

    An example of the prompt is provided in Figure ~\ref{fig:exampleofprompt}. We label each element within the prompt for clarification, though these labels should be omitted in the final version.

From this response, we can see that LLM believes the purpose of this program might be to read integers n, m, h, and an array s of size m, then performs probability calculations based on these values, producing different outputs according to their relationships. LLM localizes line 16 as a potential issue, placing it at the highest level of suspiciousness value. The reason is that the loop's termination condition should be \texttt{j <= n}
instead of 
\texttt{j < n}
, because the loop needs to include the nth element, which the former includes but the latter does not.
This example demonstrates that LLM possesses sufficient capability to understand the meaning of code and can logically localize faults.

Moreover, during our experiments, we find that LLMs occasionally have misunderstandings about the line numbers in code, as illustrated by the following brief code example:

\begin{myleftline}
\footnotesize
\textbf{Codeflaws v162:}
\begin{code}
1 #include<stdio.h>
2 int main(int argc, char *argv[])
3 {
4
5 //    freopen("x.txt","w",stdout);  ...
\end{code}
\end{myleftline}

In this code segment, the LLMs may overlook elements like headers like line 1, comments like line 5, and empty lines like line 4, that are unrelated to the logic of the code but can affect line numbers, leading to situations where LLMs might localize the correct fault location, but the outputted line number could be displaced. Therefore, we processed the code inputted to the LLMs by adding a number in front of each line of code as its line number, facilitating accurate localization by the LLMs.

To better understand the influence of each component in prompt on the fault localization performance of LLMs for novice programs, we perform ablation experiments as outlined in RQ3.

\noindent \textbf{3) Code Context}
\label{sec:Codecontext}


Current LLMs typically impose a token limit on input prompts~\cite{xue2024repeat}. If a prompt exceeds this token limit, it cannot be processed. For example, closed-source LLMs such as GPT-3.5-Turbo and GPT-4 support prompt lengths of 4,096 and 8,192 tokens, respectively. Meanwhile, open-source LLMs used in this study, including ChatGLM3, Llama2, and Code Llama, have a token limit of approximately 2,048 tokens.

Due to the significant variation in code length across datasets, a program with excessive length may exceed the token limit for the prompt, preventing it from being processed by the LLMs. To address this, we apply a selection criterion ensuring that the total token count for each prompt remains under 2,048. As a result, 503 programs from both the Codeflaws and Condefects datasets are excluded, ensuring that the remaining $503\times3$ programs can be processed by the LLMs.


To ensure fairness in the experiments, we adopt these 1,509 programs as the unified dataset for all comparison methods in the subsequent analyses. The list of selected programs is publicly available in our code repository, supporting the replication and verification of the experimental results.

\noindent \textbf{4) Collection of Results}

To streamline the batch processing of experimental results, we require LLMs to output fault localization results in a standardized JSON format. Given that LLMs may include extraneous non-JSON text, we use a predefined regular expression to extract the necessary JSON data. ~\revise{ The regular expressions are as follows.} If the output fails to meet the specified standards, we resubmit the prompt until a valid response is obtained. \revise{ In order to ensure the effective output and system cost, we set the upper limit of repetition to 10 times, but in experiments we find that the number of repetitions is mostly within 3 times}

\begin{myleftline}
\footnotesize
\textbf{regular expression to process the result :}
\begin{code}
 pattern = r'\{.{0,4000}"faultLocalization".*\[.*\].{0,4000}\}'
\end{code}
\end{myleftline}

After receiving the JSON output from the LLMs, we assess their performance using the Top-N metric by matching the results in the ``faultLocalization" array with the actual fault locations. This approach streamlines data processing and offers a precise metric for evaluating LLM efficiency in novice program fault localization tasks.

\noindent \textbf{5) Questionnaire Setup}

\revise{To evaluate the plausibility of fault explanations generated by LLM, we designed a questionnaire-based study following the methodology of Ng et al.~\cite{ng2024design}.  The questionnaire samples are drawn from 30 programs on the BugT dataset, where both closed-source LLMs and open-source LLMs successfully perform fault localization. Each questionnaire include the fault version, its corresponding correct version, the problem description, test cases, and the model-generated explanation of the fault. We recruit 10 novice programmers as participants, consisting of five individuals with one year of programming experience and five with three years of experience. During the analysis phase, we conduct a comparative and consistency analysis between the two participant groups to enhance the rigor and reliability of the results.}

\section{Research Questions}
\label{sec:Research Question}

To conduct a thorough and comprehensive validation of the fault localization capabilities of LLMs on novice code, we propose the following research questions. These questions aim to provide a detailed analysis, ranging from the design of prompts for LLMs to comparative analysis of the model’s results. 
\begin{itemize}
  \item 
\textbf{RQ1: How do different LLMs perform when localizing faults on novice programs?}

This research question examines the fault localization performance of different LLMs compared to traditional fault localization techniques, like SBFL and MBFL, in helping novice programmers identify faults. It aims to evaluate the potential of LLMs for improving fault localization accuracy specifically for novice programmers.

   \item 
\textbf{How do the fault localization results of LLM-based and traditional techniques compare in terms of unique fault localization?}  
  
Different fault localization methods perform differently in different programs. To investigate the correlation and complementary between different methods, this RQ collects the overlap in faults identified by different techniques, as well as the unique and missed faults for each approach. The aim is to uncover the relative strengths and limitations of these approaches through this analysis.

  \item \textbf{RQ3: How does prompt engineering affect fault localization performance ccross different LLMs?}

This research question evaluates how different prompt elements influence LLM performance in fault localization for novice programmers, using ablation experiments. By systematically removing key prompt elements and measuring changes in accuracy, we aim to quantify the contribution of each component to LLM performance. This analysis seeks to identify the essential components of an effective prompt and their role in improving LLM fault localization capabilities.
    
    

    \item  \textbf{RQ4: How accuracy of LLM fault localization is across different difficulty levels of the tested programming problems?}

    This study investigates the accuracy of LLM fault localization across different levels of difficulty in the tested programming problems. Using the Codeflaws, Condefects, and BugT datasets, we evaluate model performance using the metric Top-N. By categorizing the problems into various difficulty ranges, we evaluate how effectively these models identify faults in both simpler and more complex scenarios.

    \item \textbf{RQ5: How do the explanations generated by LLMs perform in terms of quality for novice fault localization?}
   
    
This study evaluates the quality of explanations in LLM fault localization. We assess the correctness of these explanations using LLMs and conduct a survey to collect statistical evaluations of the explanation quality. This research question quantifies the quality of incorrect explanations in LLM fault localization.

\end{itemize}

\revise{\section{Experiment Results and Analysis}}
\label{sec:Result Analysis}

\subsection{RQ1: How do different LLMs perform when localizing faults on novice programs?}


In this RQ, we compare \revise{13 LLMs (six closed-source and seven open-source)} against two traditional fault localization methods (SBFL and MBFL). The fault localization performance is evaluated using the widely adopted Top-N metric, where a higher Top-N value indicates the fault localization technique has a better performance. The experimental results of different techniques on three datasets are presented in Table~\ref{tab:accurancy}. Take the number 99 in the first cell of the first row in Table~\ref{tab:accurancy} (a) as an example, it indicates that using OpenAI o3 can localize 99 faults at the first position in the suspiciousness list. We highlight the best-performing method in each column in blue color.

\begin{table}[!htbp]
 \renewcommand\arraystretch{0.93} 
\centering
\caption{Effectiveness Comparison of LLM-based FL and Traditional FL}
 \vspace{-4mm}
 \begin{subtable}{0.48\textwidth}
\centering
\caption{Codeflaws}
\resizebox{\textwidth}{!}{
    \begin{tabular}{cclccccc}
    \toprule
    \multicolumn{3}{c}{\textbf{Technique}} & \textbf{Top-1} & \textbf{Top-2} & \textbf{Top-3} & \textbf{Top-4} & \textbf{Top-5} \\
\midrule
\multirow{13}[0]{*}{\begin{sideways}\textbf{LLM-based FL}\end{sideways}} & \multirow{5}[0]{*}{\begin{sideways}\textbf{Closed-Source}\end{sideways}} 
& \textbf{OpenAI o3} & 99    & 153   & 215 \cellcolor[HTML]{C4D8F2}  & 257 \cellcolor[HTML]{C4D8F2}  & 291\cellcolor[HTML]{C4D8F2} \\
        &       & \textbf{o1-preview} & 90    & 136   & 180   & 219   & 253 \\
          &       & \textbf{o1-mini} & 83    & 141   & 182   & 232   & 266 \\
                  &       & \textbf{GPT-4o} & 101   & 166   & 213    & 257    & 287 \\
          &       & \textbf{GPT-4} & 90    & 152   & 198   & 235   & 266 \\

          &       & \textbf{GPT-3.5-Turbo} & 100   & 157   & 194   & 221   & 240 \\
\cmidrule{2-8}          & \multirow{7}[2]{*}{\begin{sideways}\textbf{Open-Source}\end{sideways}} & \textbf{ChatGLM4} & 52    & 94    & 132   & 163   & 197 \\
          &       & \textbf{ChatGLM3} & 9     & 28    & 54    & 82    & 106 \\
          &       & \textbf{\revise{DeepSeekR1}} & 110\cellcolor[HTML]{C4D8F2}   & 169   & 206   & 238   & 280  \\
          &       & \textbf{\revise{DeepSeekV3}} & 107    & 173 \cellcolor[HTML]{C4D8F2}   & 212   & 247   & 291  \\
          &       & \textbf{Llama3} & 58    & 102   & 132   & 160   & 178 \\
          &       & \textbf{Llama2} & 8     & 27    & 49    & 68    & 89 \\
          &       & \textbf{Code Llama} & 41    & 73    & 104   & 126   & 136 \\
    \midrule
    \multirow{6}[4]{*}{\begin{sideways}\textbf{Tranditional FL}\end{sideways}} & \multirow{3}[2]{*}{\begin{sideways}\textbf{SBFL}\end{sideways}} & \textbf{Dstar} & 13    & 34    & 54    & 75    & 106 \\
          &       & \textbf{Ochiai} & 13    & 34    & 54    & 75    & 106 \\
          &       & \textbf{Op2} & 13    & 34    & 54    & 75    & 107 \\
\cmidrule{2-8}          & \multirow{3}[2]{*}{\begin{sideways}\textbf{MBFL}\end{sideways}} & \textbf{Dstar} & 3     & 8     & 24    & 37    & 75 \\
          &       & \textbf{Ochiai} & 3     & 8     & 24    & 37    & 75 \\
          &       & \textbf{Op2} & 3     & 8     & 24    & 37    & 75 \\
    \bottomrule
    \end{tabular}%
}
\label{subtable:1}
\end{subtable}%

\begin{subtable}{0.48\textwidth}
\centering
\caption{Condefects}
\resizebox{\textwidth}{!}{
    \begin{tabular}{cclccccc}
    \toprule
    \multicolumn{3}{c}{\textbf{Technique}} & \textbf{Top-1} & \textbf{Top-2} & \textbf{Top-3} & \textbf{Top-4} & \textbf{Top-5} \\
\midrule
\multirow{12}[0]{*}{\begin{sideways}\textbf{LLM-based FL}\end{sideways}} & \multirow{5}[0]{*}{\begin{sideways}\textbf{Closed-Source}\end{sideways}} & \textbf{OpenAI o3} & 312\cellcolor[HTML]{C4D8F2}   & 357\cellcolor[HTML]{C4D8F2}   & 373\cellcolor[HTML]{C4D8F2}   & 394\cellcolor[HTML]{C4D8F2}   & 409\cellcolor[HTML]{C4D8F2} \\
        &       & \textbf{o1-preview} & 258   & 312   & 344   & 366   & 378 \\
          &       & \textbf{o1-mini} & 211   & 271   & 306   & 333   & 348 \\
          
        &       & \textbf{GPT-4o} & 213   & 263   & 310   & 331   & 360 \\
          &       & \textbf{GPT-4} & 200   & 261   & 300   & 332   & 348 \\

          &       & \textbf{GPT-3.5-Turbo} & 148   & 212   & 248   & 274   & 290 \\
\cmidrule{2-8}          & \multirow{7}[2]{*}{\begin{sideways}\textbf{Open-Source}\end{sideways}} & \textbf{ChatGLM4} & 78    & 152   & 206   & 240   & 273 \\
          &       & \textbf{ChatGLM3} & 22    & 51    & 90    & 126   & 152 \\
          &       & \textbf{\revise{DeepSeekR1}} & 290   & 340   & 371   & 389   & 401 \\
          &       & \textbf{\revise{DeepSeekV3}} & 215   & 283   & 316   & 328   & 345 \\
          &       & \textbf{Llama3} & 100   & 168   & 209   & 231   & 251 \\
          &       & \textbf{Llama2} & 19    & 43    & 66    & 101   & 131 \\
          &       & \textbf{Code Llama} & 59    & 112   & 147   & 169   & 180 \\
    \midrule
    \multirow{6}[4]{*}{\begin{sideways}\textbf{Tranditional FL}\end{sideways}} & \multirow{3}[2]{*}{\begin{sideways}\textbf{SBFL}\end{sideways}} & \textbf{Dstar} & 0     & 2     & 59    & 132   & 170 \\
          &       & \textbf{Ochiai} & 0     & 2     & 59    & 132   & 170 \\
          &       & \textbf{Op2} & 0     & 2     & 59    & 132   & 170 \\
\cmidrule{2-8}          & \multirow{3}[2]{*}{\begin{sideways}\textbf{MBFL}\end{sideways}} & \textbf{Dstar} & 84    & 172   & 228   & 262   & 300 \\
          &       & \textbf{Ochiai} & 84    & 172   & 228   & 262   & 300 \\
          &       & \textbf{Op2} & 84    & 172   & 229   & 263   & 301 \\
    \bottomrule
    \end{tabular}%
}
\label{subtable:2}
\end{subtable}%

\begin{subtable}{0.48\textwidth}
\centering
\caption{BugT}
\resizebox{\textwidth}{!}{
    \begin{tabular}{cclccccc}
    \toprule
    \multicolumn{3}{c}{\textbf{Technique}} & \textbf{Top-1} & \textbf{Top-2} & \textbf{Top-3} & \textbf{Top-4} & \textbf{Top-5} \\
\midrule
\multirow{12}[0]{*}{\begin{sideways}\textbf{LLM-based FL}\end{sideways}} & \multirow{5}[0]{*}{\begin{sideways}\textbf{Closed-Source}\end{sideways}} & \textbf{OpenAI o3} & 289\cellcolor[HTML]{C4D8F2}  & 374\cellcolor[HTML]{C4D8F2}  & 413\cellcolor[HTML]{C4D8F2}  & 437\cellcolor[HTML]{C4D8F2}  & 449\cellcolor[HTML]{C4D8F2}   \\

 &       & \textbf{o1-preview} & 242  & 313  & 365  & 393  & 413   \\
          &       & \textbf{o1-mini} & 204   & 283   & 329   & 369   & 407 \\
     &       & \textbf{GPT-4o} & 177   & 240   & 288   & 321   & 358 \\
          &       & \textbf{GPT-4} & 192   & 242   & 294   & 332   & 361 \\
  
          &       & \textbf{GPT-3.5-Turbo} & 133   & 185   & 223   & 246   & 278 \\
\cmidrule{2-8}          & \multirow{7}[2]{*}{\begin{sideways}\textbf{Open-Source}\end{sideways}} & \textbf{ChatGLM4} & 111   & 150   & 182   & 219   & 247 \\
          &       & \textbf{ChatGLM3} & 34    & 63    & 81    & 107   & 130 \\
          &       & \textbf{\revise{DeepSeekR1}} & 248   & 308   & 345   & 380   & 424   \\
          &       & \textbf{\revise{DeepSeekV3}} & 209   & 258   & 294   & 326   & 366 \\
          &       & \textbf{Llama3} & 102   & 149   & 182   & 225   & 253 \\
          &       & \textbf{Llama2} & 40    & 65    & 102   & 124   & 147 \\
          &       & \textbf{Code Llama} & 49    & 78    & 108   & 124   & 145 \\
    \midrule
    \multirow{6}[4]{*}{\begin{sideways}\textbf{Tranditional FL}\end{sideways}} & \multirow{3}[2]{*}{\begin{sideways}\textbf{SBFL}\end{sideways}} & \textbf{Dstar} & 1     & 8     & 15    & 17    & 20 \\
          &       & \textbf{Ochiai} & 1     & 8     & 15    & 17    & 20 \\
          &       & \textbf{Op2} & 1     & 8     & 15    & 17    & 20 \\
\cmidrule{2-8}        

& \multirow{3}[2]{*}{\begin{sideways}\textbf{MBFL}\end{sideways}} &
\textbf{Dstar} & 2     & 16    & 46    & 108   & 135 \\
          &       & \textbf{Ochiai} & 2     & 16    & 46    & 108   & 135 \\
          &       & \textbf{Op2} & 2     & 16    & 46    & 108   & 135 \\
    \bottomrule
    \end{tabular}%
}
\label{subtable:3}
\end{subtable}%

\label{tab:accurancy}
\end{table}

The results in Table~\ref{tab:accurancy} demonstrate that both closed-source and open-source LLMs generally outperform traditional fault localization methods (SBFL and MBFL) across all three datasets. This suggests that LLMs are more effective at capturing complex fault patterns compared to traditional statistical-based techniques. 


In the comparison between closed-source and open-source LLMs, the results reveal a clear trend: closed-source LLMs generally achieve better fault localization performance than open-source LLMs in the Condefects and BugT datasets. Notably, \revise{OpenAI o3} consistently outperforms all other models in these two datasets, indicating its strong capability in fault localization. However, an exceptional case appears in the Codeflaws dataset, where GPT-4o surpasses all closed-source LLMs in terms of Top-1 and Top-2 accuracy. \revise{This is counterintuitive, as \revise{OpenAI o3} were released after GPT-4o, and according to the official technical documentation~cite{~\cite{openai2025o3}}, \revise{OpenAI o3} have stronger performance. We will further discuss this situation in Section}~\ref{subsec:discuss-paradox}.


\revise{Among the seven open-source LLMs}, \revise{DeepSeekR1} stands out as the best-performing model in fault localization. This superiority is particularly evident in the Codeflaws dataset, where DeepSeekR1 achieves the highest accuracy for both Top-1 and Top-2 metrics. A possible reason for this advantage is that DeepSeekR1 has been specifically trained on code-related tasks, such as code analysis and code completion~\cite{deepseekai2025deepseekv3technicalreport}, which enhances its ability to understand and locate faults. Additionally, DeepSeekR1 is the largest model in terms of parameter size among all the open-source LLMs included in our study. These findings suggest that both specialized training on code-related tasks and larger model sizes contribute to improved performance in program fault localization.

\begin{center}
\begin{tcolorbox}[colback=blue!1,
                  colframe=black,
                  arc=1mm, auto outer arc,
                  boxrule=0.5pt
                 ]
\textbf{Summary for RQ1:} 
\revise{All six closed-source LLMs perform significantly better than traditional fault localization techniques, SBFL and MBFL. Regarding seven open-source LLMs, the latest LLM, DeepSeekR1, which has been specialized and trained for code-related tasks, performs better than most closed-source LLMs and other open-source LLMs.}
\end{tcolorbox}
\end{center}

\subsection{RQ2: How do the fault localization results of LLM-based and traditional techniques compare in terms of unique fault localization?}



\begin{figure}[htbp]
    \centering
    \begin{subfigure}[b]{\linewidth}
        \hspace{-5mm}
        \includegraphics[width=1.1\linewidth]{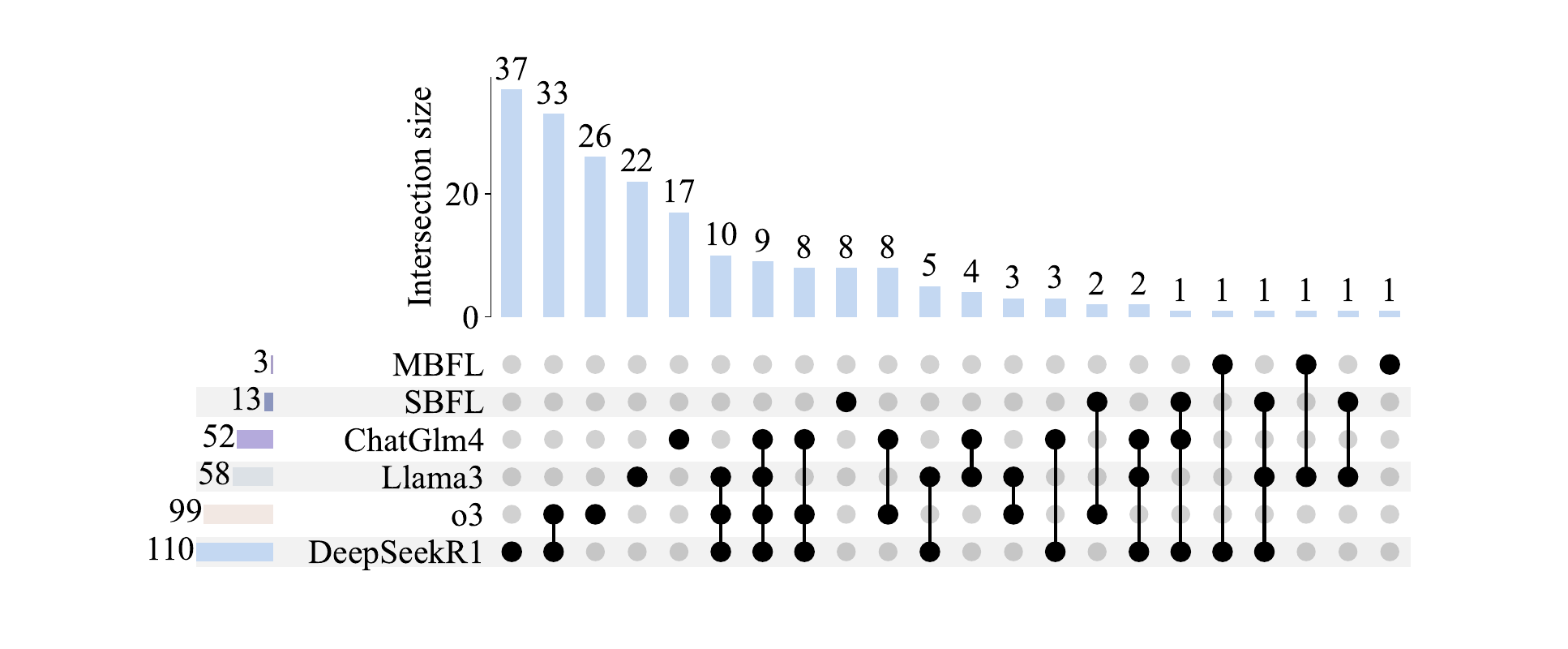}
        \caption{Codeflaws Dataset}
        \label{fig:in Codeflaws}
        \vspace{2mm}
    \end{subfigure}
    
    \begin{subfigure}[b]{\linewidth}
        \hspace{-5mm}
        \includegraphics[width=1.1\linewidth]{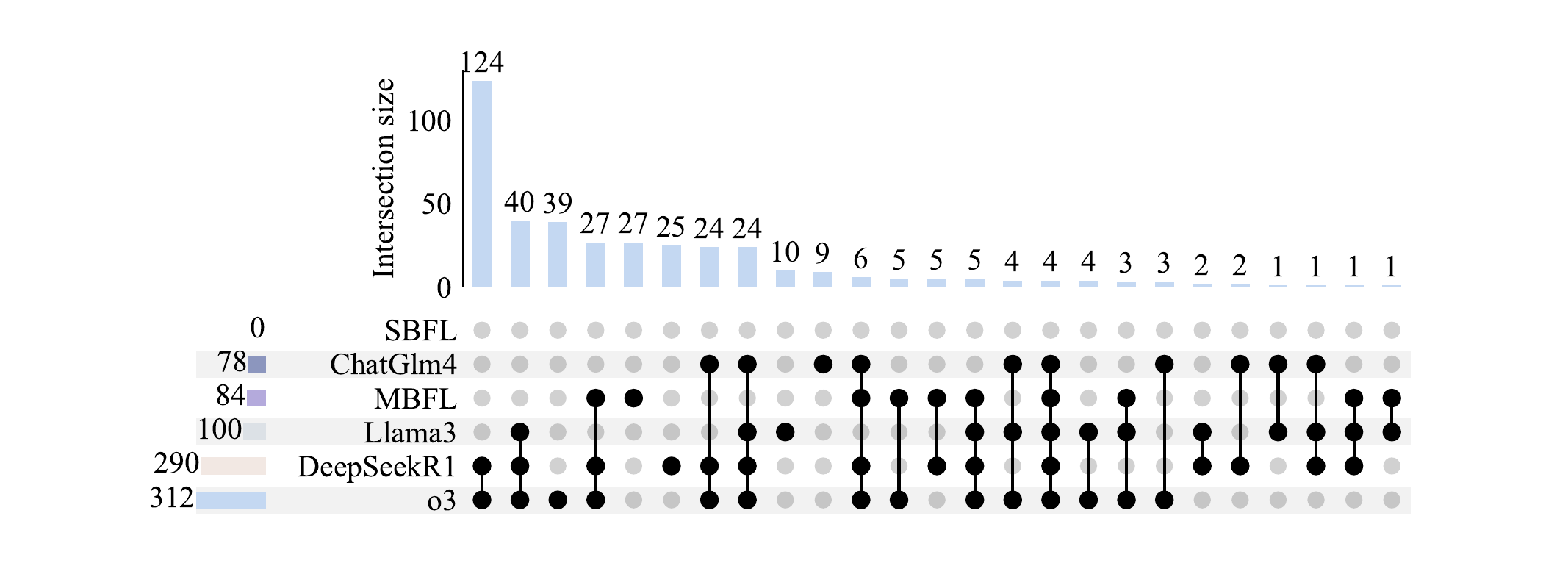}
        \caption{Condefects Dataset}
        \label{fig:in Condefects}
        \vspace{2mm}
    \end{subfigure}
    
    \begin{subfigure}[b]{\linewidth}
        \hspace{-5mm}
        \includegraphics[width=1.1\linewidth]{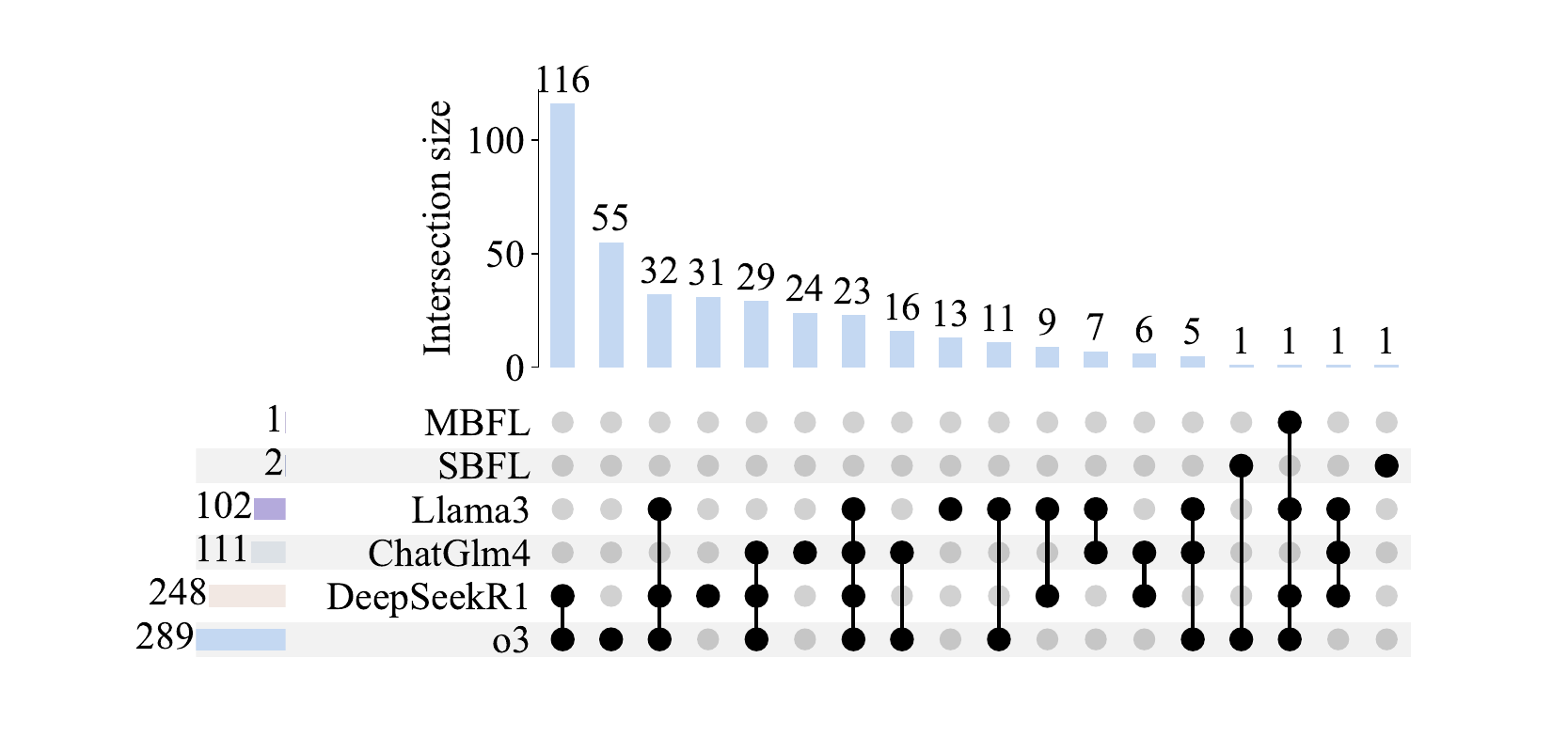}
        \caption{BugT Dataset}
        \label{fig:in BugT}
    \end{subfigure}
    
    \caption{Overlap among LLMs fault localization and Traditional techniques in different datasets}
    \label{fig:Overlap among LLMs fault localization and Traditional techniques}
\end{figure}

In this RQ, we focus on analyzing the overlap between the fault localization results of different models, helping us understand the relationships and complementarities among various models. Based on the results from RQ1, \revise{OpenAI o3}, which performs best among closed-source LLMs, and DeepSeekR1, the top-performing open-source LLM, are selected for further examination. 
We conduct a \revise{detailed analysis of the intersection patterns using UpSet plots} including LLM-based fault localization and traditional methods like SBFL and MBFL. The Top-1 fault localization results, obtained from three datasets, are presented in Figure~\ref{fig:Overlap among LLMs fault localization and Traditional techniques}. This analysis not only reveals the unique localization capabilities of each method but also demonstrates the complementary strengths between different techniques in fault localization.

\revise{In Figure~\ref{fig:Overlap among LLMs fault localization and Traditional techniques}, we present UpSet plots to visualize the intersection patterns among different fault localization techniques. In these plots, the bottom horizontal bar chart shows the total number of faults localized by each individual method, the connected dots with lines indicate specific intersection combinations, and the top vertical bar chart displays the size of each intersection. The following is the detailed analysis from the LLM-based fault localization perspective and the traditional fault localization perspective separately.}

As illustrated in Figure~\ref{fig:Overlap among LLMs fault localization and Traditional techniques}, OpenAI o3 and DeepSeekR1 exhibit higher unique fault localization performance compared to other models.
\revise{Specifically, in the Codeflaws dataset, OpenAI o3 uniquely localized 26 faults, while DeepSeekR1 uniquely localized 37 faults. This pattern is consistent across other datasets as well.} The reason for this superior performance is that OpenAI o3 and DeepSeekR1, due to their architecture and extensive training data, possess stronger reasoning, generalization, and fault localization capabilities, which lead to their significantly better unique fault localization ability compared to other models.




\revise{In contrast, traditional fault localization techniques demonstrate significantly lower unique localization capabilities. Specifically, in the Codeflaws dataset, MBFL uniquely localized only 1 fault, while SBFL uniquely localized 8 faults.}
 This pattern is consistent across other datasets, revealing the limitations of traditional methods in independent fault localization. The primary reason lies in the different fault localization fundamentals. SBFL relies on test case coverage information to compute suspiciousness values, making it incapable of analyzing deeper semantic relationships within the code. MBFL attempts to improve fault localization through mutation-based perturbations, yet its effectiveness is constrained by predefined mutation rules, making it less suitable for capturing complex, context-dependent faults. 
In contrast, LLM-based fault localization methods do not require test execution and instead leverage semantic reasoning to identify faults, allowing them to localize a broader range of issues with greater independence.

\begin{center}
\begin{tcolorbox}[colback=blue!1,
                  colframe=black,
                  arc=1mm, auto outer arc,
                  boxrule=0.5pt
                 ]
\textbf{Summary for RQ2:} 
OpenAI o3 and DeepSeekR1 demonstrate higher unique fault localization performance compared to smaller size models and traditional methods such as SBFL and MBFL, consistently showing better results across various datasets.
\end{tcolorbox}
\end{center}

\begin{figure*}[htp]
    \centering
    
    \begin{subfigure}[b]{0.7\textwidth}
    \centering
    \includegraphics[width=\textwidth]{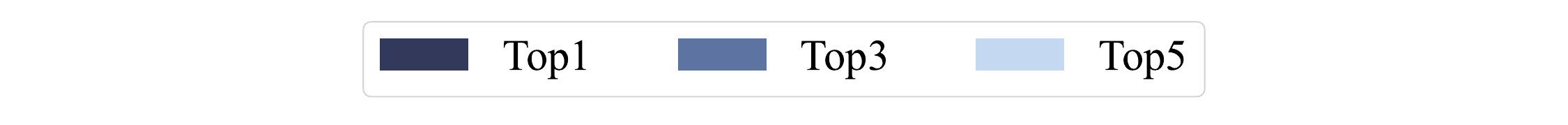}
    \vspace{-2mm}
    \end{subfigure}
    \begin{subfigure}[b]{1\textwidth}
    \centering
    \includegraphics[width=\textwidth]{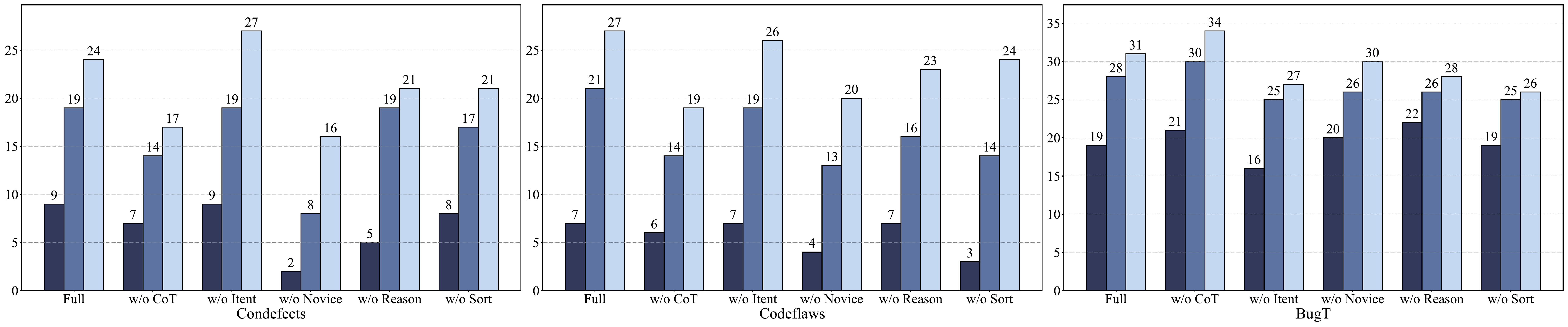}
    \vspace{-2mm}
    \caption{OpenAI o3}
    \label{fig:Condefects Prompt Various inDeepSeekV3}

    \end{subfigure}
    
    \begin{subfigure}[b]{1\textwidth}
    \centering
    \includegraphics[width=\textwidth]{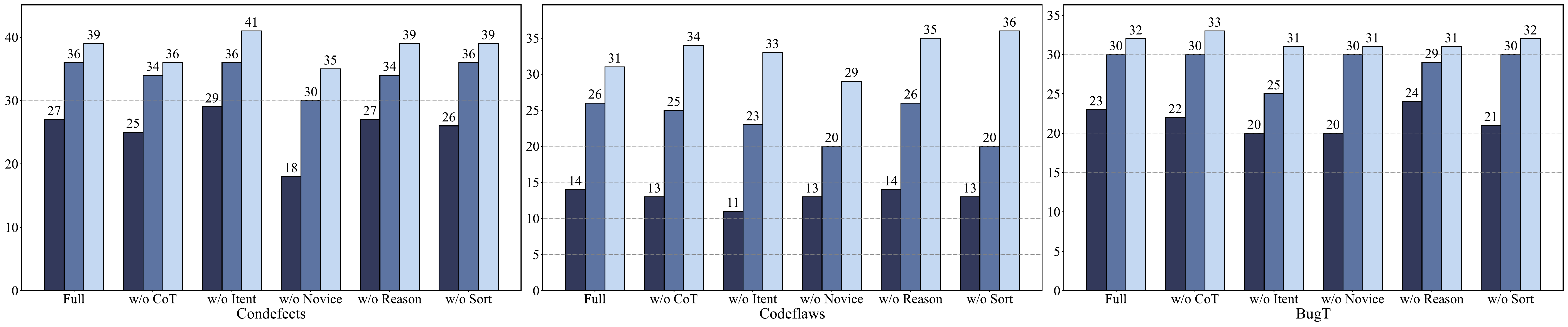}
    \vspace{-2mm}
    \caption{DeepSeekR1}
    \label{fig:BugT Prompt Various in DeepSeekV3}
    \end{subfigure}
    \begin{subfigure}[b]{1\textwidth}
        \centering
        \includegraphics[width=\textwidth]{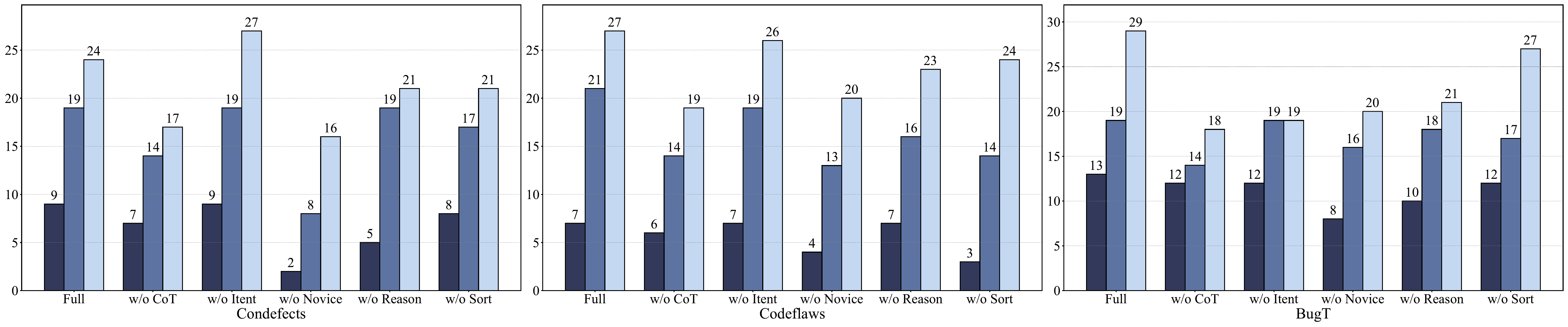}
        \vspace{-2mm}
        \caption{GPT-4}
        \label{fig:Codeflaws Prompt Various in DeepSeekV3}
    \end{subfigure}

    \centering


    \caption{Ablation Results of Prompt Components on OpenAI o3, DeepSeekR1, and GPT-4}
    \label{fig:Prompt Various in DeepSeekV3}
\end{figure*}

\subsection{RQ3: How does prompt engineering affect fault localization performance ccross different LLMs?}





In this RQ, we focus on prompt engineering and conduct an ablation study on \revise{OpenAI o3}, DeepSeekR1, and GPT-4. The selection of OpenAI o3 and DeepSeekR1 is based on the findings from RQ1, where OpenAI o3 demonstrated the best performance in fault localization, while DeepSeekR1 excelled in fault localization within open-source LLMs. In addition, GPT-4 is included to evaluate the performance of an LLM that lacks reasoning capabilities, providing a contrasting perspective to the other models. Due to the practical cost constraints of utilizing the OpenAI o3 API, we randomly select 50 program samples from each dataset—Codeflaws, Condefects, and BugT. The goal is to assess the contribution of each element in our prompt to the overall effectiveness of novice program fault localization.


Specifically, for our complete prompt design, we systematically remove different elements to create various ablated prompts. These ablated prompts are then input into  \revise{OpenAI o3}, DeepSeekR1, and GPT-4 to compare the fault localization results for novice programs. Within the comprehensive prompt, we identify five crucial elements: (1) the background description of novice programs, (2) the use of CoT to address the fault localization problem, (3) the requirement for the model to describe its understanding of the code's intent in the results, (4) the need for the model to justify why a particular line is considered the fault location, (5) the requirement for the model to rank the list of faults in descending order of suspiciousness. By excluding each of these elements one by one, we generate five ablated prompt variants, which allow us to assess the importance and contribution of each element in the effectiveness of LLM-based fault localization on novice programs:

\begin{itemize}
  \item 
  w/o CoT: This variant removes the expression of the Chain of Thought from the prompt.
  \item 
  w/o Intent: This variant removes the requirement from the prompt for LLMs to describe the intent of the code in their response.
    \item 
  w/o Novice: This variant removes the novice description of the background information about novice programs that we design from the prompt.
    \item 
  w/o Reason: This variant removes the requirement from the prompt to return the reason why it considers a particular line as the fault location during novice program fault localization.
    \item 
  w/o Sort: This variant removes the requirement from the prompt to sort the returned list of faults by descending order of suspiciousness value.
\end{itemize}

\renewcommand{\dblfloatpagefraction}{.9}

Figure~\ref{fig:Prompt Various in DeepSeekV3} shows the results of the ablation study results of prompts variants on the Codeflaws, Condefects, and BugT dataset in terms of Top-1, Top-3 and Top-5.
From Figure~\ref{fig:Prompt Various in DeepSeekV3}, we can observe that OpenAI o3 and DeepSeekR1 show minimal effects in the prompt ablation experiment. In fact, when certain elements are removed from the prompts, their performance is comparable to or even surpasses the performance when those elements are included. This is in contrast to the results from GPT-4. This observation likely stems from their intrinsic capabilities: \revise{OpenAI o3's and DeepSeekR1's reasoning capabilities},  respectively reducing reliance on external prompt guidance for fault localization.

For LLMs with medium reasoning capabilities, such as GPT-4, prompt engineering has a significant impact. Among the various prompt elements, the novice description is the most crucial. Its removal leads to the most noticeable performance drop, with the Top-1 value decreasing by 7 (77.77\%), 3 (42.8\%), and 5 (38.36\%) in Condefects, Codeflaws, and BugT, respectively. Furthermore, removing any single element from the prompt results in a reduction of the fault localization performance.

\begin{center}
\begin{tcolorbox}[colback=blue!1,
                  colframe=black,
                  arc=1mm, auto outer arc,
                  boxrule=0.5pt
                 ]
\vspace{0.5mm}
\textbf{Summary to RQ3:} 
For LLMs with modular reasoning capabilities, such as OpenAI o3 and DeepSeekR1, prompt engineering shows minimal impact, with performance remaining comparable or even improving when certain elements are removed. In contrast, for models with limited modular thinking capabilities like GPT-4, prompt engineering is crucial, especially the novice program prompt, whose removal leads to significant performance drops in fault localization.
\end{tcolorbox}
\end{center}

\subsection{RQ4: How accuracy of LLM fault localization is across different difficulty levels of the tested programming problems?}

To investigate the relationship between program difficulty and fault localization accuracy, we evaluate 13 LLMs (six closed-source and seven open-source models) across three datasets categorized by difficulty levels. 
The \revise{experiment} design involved ranking samples within each dataset based on their difficulty levels and dividing them into five equally sized groups to ensure statistical reliability, each containing 100 samples. Table~\ref{tab:difficulty} presents the detailed results, where ``Lv'' indicates the average difficulty level within each group (i.e., Lv.5 = highest difficulty to Lv.1 = lowest), with Top-1 accuracy as our evaluation metric.

\begin{table}[htbp]
    \centering
        \caption{Effectiveness Comparison of LLM-based FL on Different Difficulty Levels in terms of Top-1 Metric}
  
    \begin{subtable}{0.5\textwidth}
      \centering
      \caption{Codeflaws}
    \begin{tabular}{lrrrrr}
    \toprule
    \multicolumn{1}{c}{\multirow{2}[4]{*}{\textbf{Technique}}}  & \multicolumn{5}{c}{\textbf{Difficulty Level}} \\
\cmidrule{2-6}          & \multicolumn{1}{c}{\textbf{Lv.5}} & \multicolumn{1}{c}{\textbf{Lv.4}} & \multicolumn{1}{c}{\textbf{Lv.3}} & \multicolumn{1}{c}{\textbf{Lv.2}} & \multicolumn{1}{c}{\textbf{Lv.1}} \\
    \midrule
    \textbf{OpenAI o3} & 44    & 55    & 63    & 65    & 64 \\
    \textbf{o1-preview} & 46    & 43    & 51    & 51    & 61 \\
    \textbf{o1-mini} & 43    & 49    & 54    & 58    & 60 \\
    \textbf{GPT-4o} & 42    & 55    & 63    & 65    & 61 \\
    \textbf{GPT-4} & 35    & 55    & 55    & 54    & 65 \\
    \textbf{GPT-3.5-Turbo} & 27    & 42    & 56    & 57    & 55 \\
    \textbf{ChatGLM4} & 32    & 43    & 36    & 41    & 45 \\
    \textbf{ChatGLM3} & 17    & 27    & 19    & 23    & 20 \\
    \textbf{\revise{DeepSeekR1}} & 45    & 58    & 57    & 61    & 58 \\
    \textbf{\revise{DeepSeekV3}} & 49    & 53    & 57    & 66    & 64 \\
    \textbf{Llama3} & 24    & 32    & 36    & 40    & 46 \\
    \textbf{Llama2} & 16    & 12    & 26    & 20    & 15 \\
    \textbf{Code Llama} & 18    & 28    & 28    & 35    & 27 \\
    \bottomrule
    \end{tabular}%
      \label{tab:Codeflaws}%
    \end{subtable}%

    \begin{subtable}{0.48\textwidth}
      \centering
      \caption{Condefects}
    \begin{tabular}{lccccc}
    \toprule
    \multicolumn{1}{c}{\multirow{2}[4]{*}{\textbf{Technique}}}  & \multicolumn{5}{c}{\textbf{Difficulty Level}} \\
\cmidrule{2-6}          & \textbf{Lv.5} & \textbf{Lv.4} & \textbf{Lv.3} & \textbf{Lv.2} & \textbf{Lv.1} \\
    \midrule
    \textbf{OpenAI o3} & 71    & 74    & 86    & 85    & 91 \\
    \textbf{o1-preview} & 57    & 65    & 85    & 83    & 86 \\
    \textbf{o1-mini} & 48    & 57    & 79    & 76    & 86 \\
    \textbf{GPT-4o} & 54    & 60    & 77    & 80    & 88 \\
    \textbf{GPT-4} & 56    & 53    & 75    & 73    & 89 \\
    \textbf{GPT-3.5-Turbo} & 37    & 46    & 59    & 72    & 75 \\
    \textbf{ChatGLM4} & 22    & 36    & 56    & 73    & 85 \\
    \textbf{ChatGLM3} & 21    & 20    & 24    & 37    & 50 \\
    \textbf{\revise{DeepSeekR1}} & 66    & 77    & 81    & 83    & 92 \\
    \textbf{\revise{DeepSeekV3}} & 47    & 54    & 79    & 79    & 85 \\
    \textbf{Llama3} & 24    & 36    & 53    & 65    & 72 \\
    \textbf{Llama2} & 15    & 19    & 25    & 32    & 40 \\
    \textbf{Code Llama} & 18    & 26    & 36    & 43    & 57 \\
    \bottomrule
    \end{tabular}%
      \label{tab:Condefects}%
    \end{subtable}%
    
    \begin{subtable}{0.48\textwidth}
      \centering
      \caption{BugT}
    \begin{tabular}{lrrrrr}
    \toprule
    \multicolumn{1}{c}{\multirow{2}[4]{*}{\textbf{Technique}}}  & \multicolumn{5}{c}{\textbf{Difficulty Level}} \\
\cmidrule{2-6}          & \multicolumn{1}{c}{\textbf{Lv.5}} & \multicolumn{1}{c}{\textbf{Lv.4}} & \multicolumn{1}{c}{\textbf{Lv.3}} & \multicolumn{1}{c}{\textbf{Lv.2}} & \multicolumn{1}{c}{\textbf{Lv.1}} \\
    \midrule
    \textbf{OpenAI o3} & 94    & 84    & 83    & 91    & 95 \\
    \textbf{o1-preview} & 89    & 74    & 76    & 82    & 90 \\
    \textbf{o1-mini} & 89    & 72    & 74    & 81    & 89 \\
    \textbf{GPT-4o} & 69    & 64    & 68    & 69    & 87 \\
    \textbf{GPT-4} & 74    & 68    & 78    & 50    & 89 \\
    \textbf{GPT-3.5-Turbo} & 40    & 48    & 72    & 53    & 63 \\
    \textbf{ChatGLM4} & 48    & 42    & 45    & 57    & 54 \\
    \textbf{ChatGLM3} & 26    & 18    & 17    & 36    & 33 \\
    \textbf{\revise{DeepSeekR1}} & 90    & 80    & 79    & 80    & 93 \\
    \textbf{\revise{DeepSeekV3}} & 75    & 68    & 73    & 68    & 81 \\
    \textbf{Llama3} & 58    & 40    & 59    & 46    & 49 \\
    \textbf{Llama2} & 23    & 20    & 31    & 41    & 31 \\
    \textbf{Code Llama} & 15    & 23    & 39    & 32    & 35 \\
    \bottomrule
    \end{tabular}%
      \label{tab:BugT}%
    \end{subtable}%
    \label{tab:difficulty}
\end{table}

From the experimental results shown in the Table~\ref{tab:difficulty}, we observe that in the Codeflaws and Condefects datasets, the Top-5 value of most models increases as the difficulty level decreases. For instance, in the Codeflaws dataset, the o1-preview model achieved a Top-5 value of 46 at Lv.5 and improved to 61 at Lv.1, representing an increase of 32.61\%. Similarly, in the Condefects dataset, its performance improved from 57 at Lv.5 to 86 at Lv.1, corresponding to a relative increase of 50.88\%.
BugT presents a contrasting pattern, where top closed-source models maintain high accuracy even at maximum difficulty. o1-preview and o1-mini achieve identical Lv.5 Top-5 (89) in BugT, outperforming their Codeflaws Lv.5 performance by 93.5\% and 106.9\% respectively. This suggests the difficulty ceiling of BugT may not sufficiently challenge SOTA closed-source models. However, performance variations emerge at lower difficulty levels: o1-preview's Top-5 score fluctuates between 74 (Lv.4) and 90 (Lv.1), indicating potential instability in handling medium-difficulty samples.

Furthermore, closed-source LLMs and DeepSeekR1 consistently demonstrate superior performance in fault localization across all datasets and difficulty levels, which is consistent with our earlier findings in RQ1. These results underscore the strengths of closed-source models through deep semantic analysis in this task while also highlighting areas where targeted optimization is needed, particularly for handling medium to high-difficulty faults.

\begin{center}
\begin{tcolorbox}[colback=blue!1,
                  colframe=black,
                  arc=1mm, auto outer arc,
                  boxrule=0.5pt
                 ]
\textbf{Summary for RQ4:} 
LLM-based fault localization shows high accuracy for simple programming problems, but this accuracy gradually decreases as problem difficulty increases. In the Codeflaws and Condefects dataset, model accuracy increases as difficulty drops. However, in the BugT dataset, top models stay accurate even at peak difficulty, hinting at its limited challenge.
\end{tcolorbox}
\end{center}


\subsection{RQ5: How do the explanations generated by LLMs perform in terms of quality for novice fault localization?}



\begin{figure*}[!htbp]
  
    \begin{subfigure}[b]{\textwidth}
        \centering
        \includegraphics[width=1\textwidth]{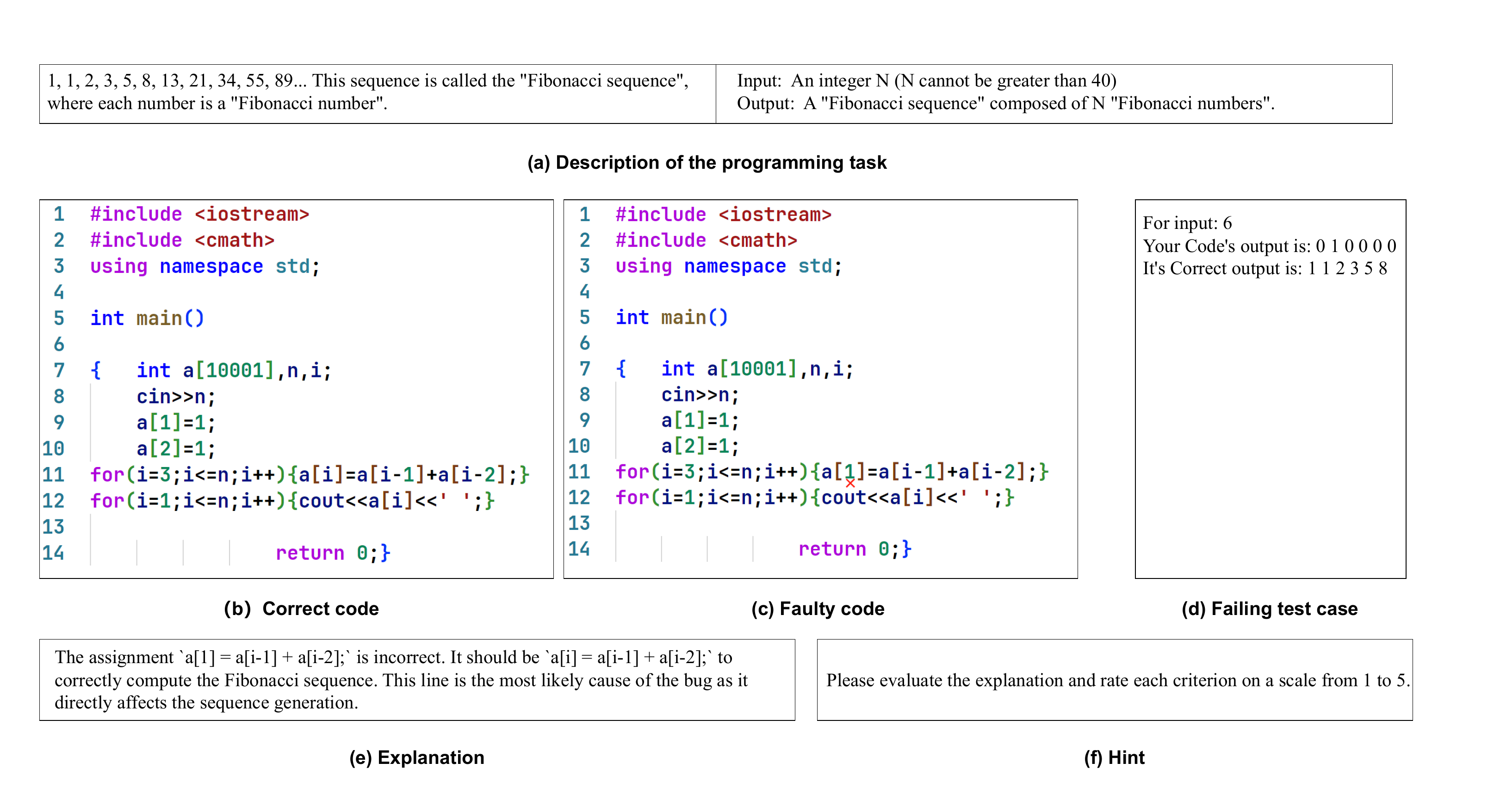}

    \end{subfigure}
  
    \caption{An Example of Questionnaire}
     \label{fig:Questionnaires example}
\end{figure*}
\revise{LLM-based programming feedback holds significant educational value, particularly for inexperienced novice programmers, where fault explanations are of paramount importance. In RQ5, we evaluate the quality and practical utility of explanations generated by LLMs in assisting novice programmers with fault localization tasks. Our objective is to determine whether these explanations effectively help novices understand and resolve faults. To accomplish this, we conduct a questionnaire survey comparing open-source and closed-source LLMs using randomly sampled examples, with evaluations performed by novice programmers across five key dimensions: readability, usefulness, conciseness, relevance, and accuracy. 
}


To conduct the manual evaluation, \revise{we respectively select 5 novice programmers with one year of programming experience and 5 programmers with three years of experience, all from the BuctOJ platform. 
Given their familiarity with the platform, we select the BugT dataset, which is collected from BuctOJ, as the research subject. 
The evaluators are divided into 2 groups based on their programming experience, each group independently evaluated 30 randomly sampled examples where the open-source and closed-source LLMs had successfully localized faults. To reduce potential subjective bias, the evaluators were not informed that the fault explanations are generated by LLMs.}
They are provided with the problem description, faulty code, correct code, failed test cases, and fault explanations automatically generated by the LLMs. The evaluation is based on the following five criteria:

\begin{itemize}
    \item 
    \textbf{Readability}: Is the explanation clear and easy to understand?   
    \item 
    \textbf{Usefulness}: Does the explanation help in understanding and fixing the fault? 
    \item 
    \textbf{Conciseness}: Does the explanation not contain unnecessary information? 
    \item 
    \textbf{Relevance}: Does the explanation directly address the fault?
    \item 
    \textbf{Accuracy}: Is the explanation technically correct?  
\end{itemize}

For better presentation, we have included one of the questionnaires in Figure~\ref{fig:Questionnaires example} and uploaded all questionnaires to our repository.

\begin{figure*}[htp]
    \centering
    \begin{subfigure}[b]{0.48\textwidth}
        \centering
        \includegraphics[width=0.95\textwidth]{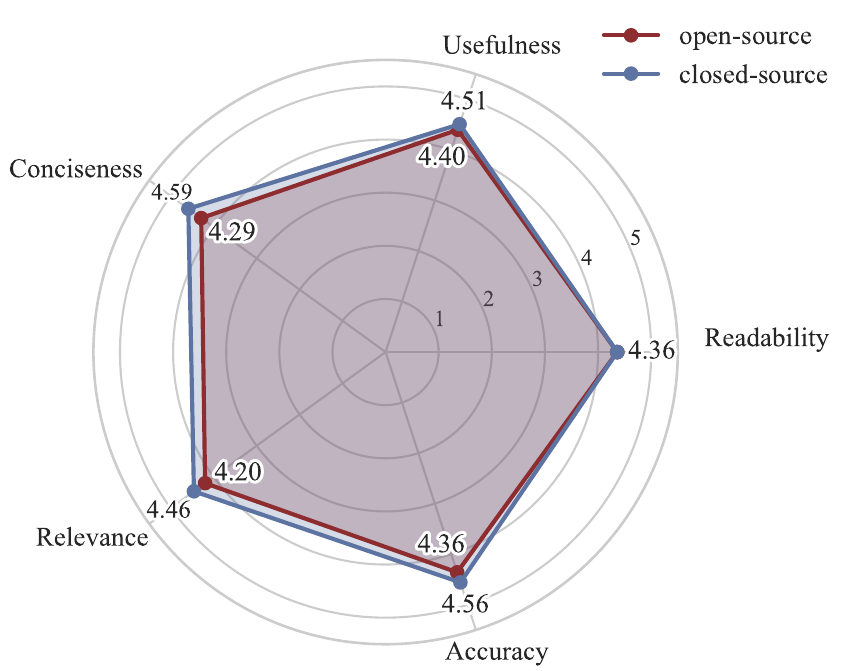}
        \caption{Comparison between LLMs}
        \label{fig:avg}
    \end{subfigure}
    \begin{subfigure}[b]{0.48\textwidth}
        \centering
        \includegraphics[width=0.95\textwidth]{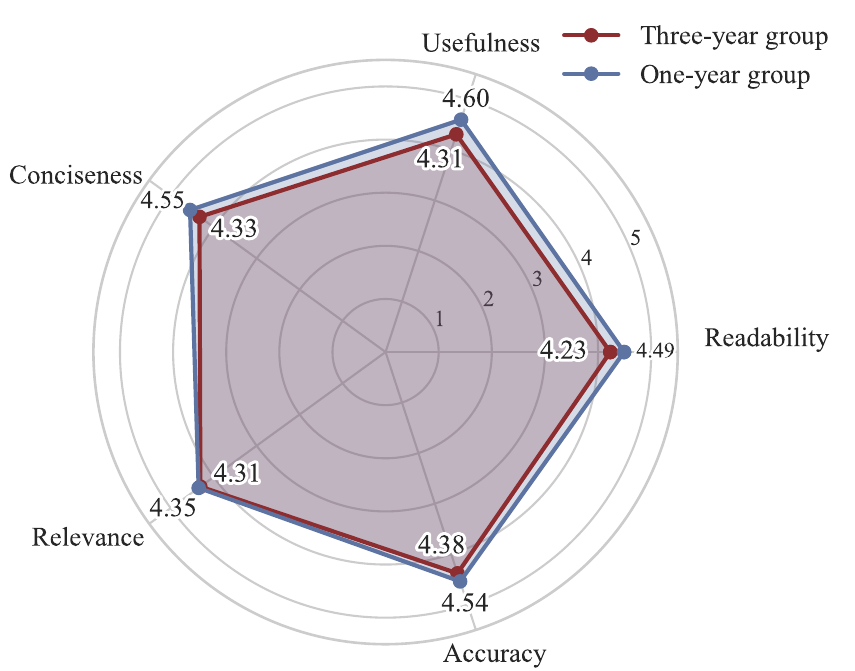}
        \caption{Comparison between experience groups}
        \label{fig:pal}
    \end{subfigure}
  
    \begin{subfigure}[b]{0.48\textwidth}
        \centering
        \includegraphics[width=0.95\textwidth]{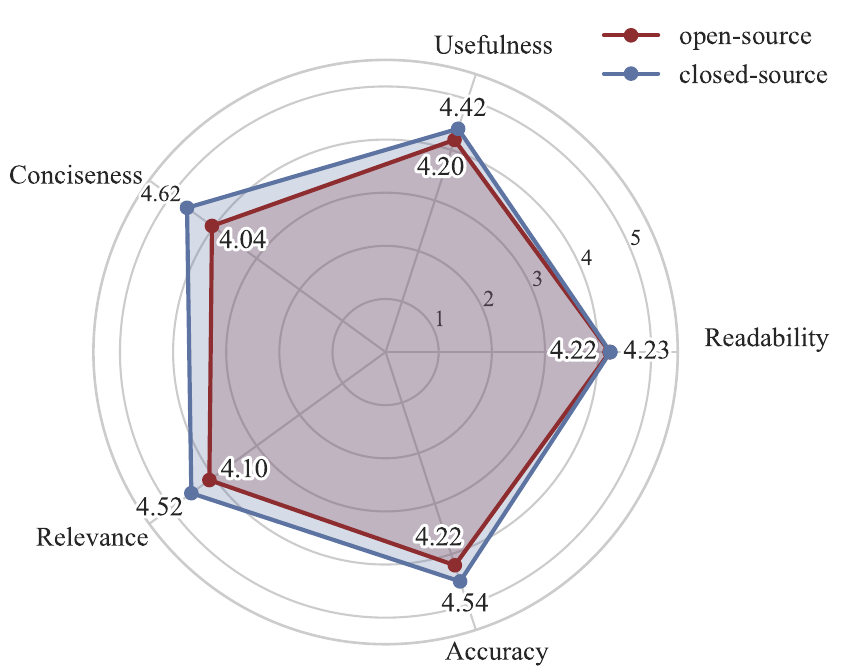}
        \caption{Three-year experience group}
        \label{fig:gosu}
    \end{subfigure}
    \begin{subfigure}[b]{0.48\textwidth}
        \centering
        \includegraphics[width=0.95\textwidth]{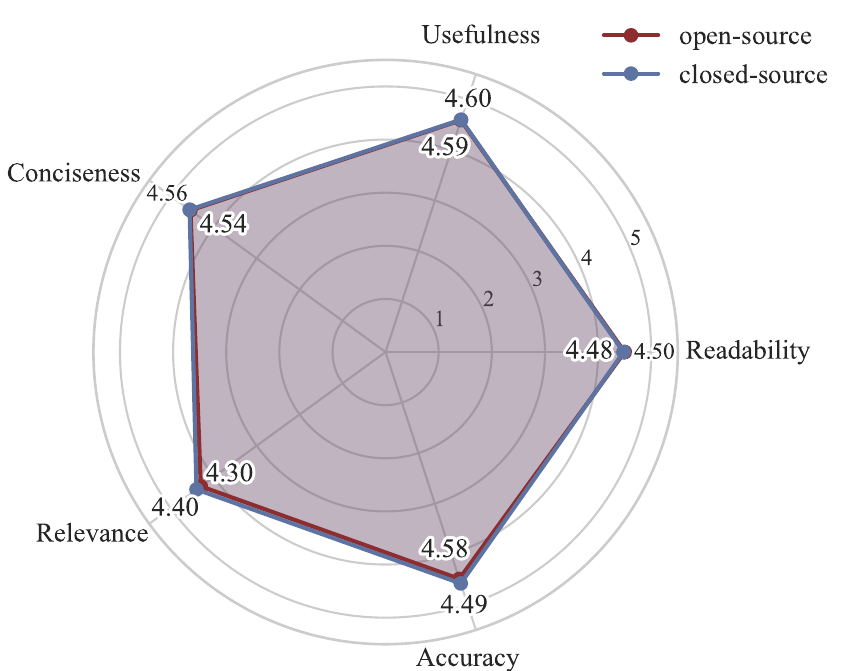}
        \caption{One-year experience group}
        \label{fig:noob}
    \end{subfigure}

    \caption{Questionnaires Survey Result}
    \label{fig:SurveyResult}
\end{figure*}

\revise{The survey results are presented in Figure~\ref{fig:SurveyResult}, where subfigure (a) displays the average scores from all participants, subfigure (b) presents a comparative analysis between the two survey groups regarding their evaluation of LLMs, and subfigures (c) and (d) illustrate the respective evaluations of different LLMs by each survey group. Our analysis proceeds in two phases: first examining the performance differences between open-source and closed-source models, followed by an investigation of the evaluation patterns across different experience groups.}

\revise{Comparison between Open-source and Closed-source Models: Regarding the comparison between open-source and closed-source LLMs, both model categories achieved identical scores of 4.36 in readability, indicating consistent explanation clarity across model types. However, closed-source models demonstrated superior performance across the remaining four evaluation dimensions. The most pronounced difference was observed in the conciseness dimension, where closed-source models outperformed open-source models by 0.30 points, representing the largest performance gap among all evaluated metrics.}

\revise{A significant finding emerged from the comparative analysis between experience groups: participants with one year of programming experience consistently provided higher ratings across all five evaluation dimensions compared to their three-year experience counterparts. This pattern suggests that programming experience significantly influences evaluation outcomes. Novice programmers with one year of experience demonstrated remarkable consistency in their assessments of both model categories, with minimal scoring variations between open-source and closed-source models. In contrast, participants with three years of programming experience exhibited more pronounced discriminatory capabilities, particularly in the conciseness and relevance dimensions.We attribute this phenomenon to the enhanced critical thinking abilities that develop with programming experience. More experienced developers appear to possess greater capacity for identifying limitations in LLM-generated explanations, consequently treating these outputs as supplementary references rather than authoritative guidance. This critical perspective enables them to provide more nuanced and differentiated evaluations.}

\revise{Particularly noteworthy is the consistently high evaluation of explanation usefulness by the one-year experience group, a finding with significant pedagogical implications. This result demonstrates that LLM-generated fault explanations possess substantial educational value for programming novices, effectively supporting their understanding of error causation and debugging skill development. These findings provide empirical support for the integration of LLMs in programming education, particularly highlighting their promising applications in novice programmer skill development and educational scaffolding.}

\begin{center}
\begin{tcolorbox}[colback=blue!1,
                  colframe=black,
                  arc=1mm, auto outer arc,
                  boxrule=0.5pt
                 ]
\textbf{Summary for RQ5:} 

\revise{Closed-source models outperform open-source models across four dimensions, with the largest gap (0.30 points) in conciseness. Programming experience significantly influences evaluations: one-year participants consistently rate explanations higher, while three-year participants show enhanced discriminatory capabilities. High usefulness ratings from novices demonstrate that LLM explanations effectively bridge knowledge gaps for programming beginners.}
\end{tcolorbox}
\end{center}

\section{Discussion}
\label{sec:Discussion}


\subsection{Advantages and disadvantages of LLMs}

Our experiments demonstrate that LLM-based FL generally outperforms traditional FL on novice programs, with traditional methods only occasionally achieving better results.  In contrast to traditional methods, which typically identify a single fault location, LLMs not only localize faults but also provide explanations and guidance, offering enhanced practical support for users~\cite{qin2024agentfl}. 
To further discuss the strengths and limitations of LLM-based FL compared to Traditional FL, we conduct detailed factor analyses and case studies, and subsequently discuss the cost implications of closed-source LLMs.

\textbf{1) Advantages of LLM}

The primary factors contributing to the superior performance of LLMs over traditional methods can be summarized as follows:




\textbf{\revise{Fault Explanation} :} The explainability advantage of LLMs in fault localization primarily lies in their ability to generate clear fault descriptions and repair suggestions through natural language. Unlike traditional fault localization tools, which rely on complex rules and static analysis results that are often challenging to interpret, LLMs can leverage contextual information to produce human-readable explanations. For instance, they can not only identify potential issues in code but also explain the root causes of the problems and provide detailed steps for resolution. This natural language-based output significantly lowers the barrier for developers to understand faults, making LLMs particularly appealing to beginners and cross-disciplinary teams. Furthermore, LLMs can infer the developer’s intent based on code comments and variable naming, offering semantic-level explanations that combine with code logic to provide deeper and more intuitive insights into complex faults. This explainability introduces a more human-centric and intelligent approach to fault localization, offering developers a more efficient and user-friendly solution.

\textbf{Chain of Thought:} The CoT capability of LLMs is a significant advantage, enabling them to gradually analyze complex problems step by step and produce more precise answers. This ability is particularly crucial in fault localization tasks, as code issues often exhibit hierarchical structures and strong contextual dependencies. Through CoT capability, LLMs can break down complex problems into smaller, manageable steps, systematically analyzing code logic, variable relationships, and potential points of conflict. Ultimately, this process results in a complete and clear fault localization pathway. This stepwise reasoning not only improves the accuracy of fault localization but also provides developers with a transparent thought process, aiding in better understanding and resolution of issues.

\textbf{Context Consideration:} Traditional fault localization methods primarily rely on code coverage information, whereas LLMs benefit from a broader context, incorporating code structure, variable naming conventions, comments, and documentation. This comprehensive understanding notably enhances the precision of fault localization.

For instance, taking program v17 from the Codeflaw as an example:

\begin{myleftline}
\footnotesize
\textbf{Codeflaws v17:}
\begin{code}
...
5     int n,a,b,c,d,i;
6     scanf("%d",&n);
(*@\highlightline@*)7     for(i=n+1;;i++)
8     {
9         a=i/1000;
10        b=(i\%1000)/100;
...
\end{code}
\end{myleftline}

The response from the LLM is as follows:

\begin{myleftline}
\footnotesize
\includegraphics[width=1\linewidth]{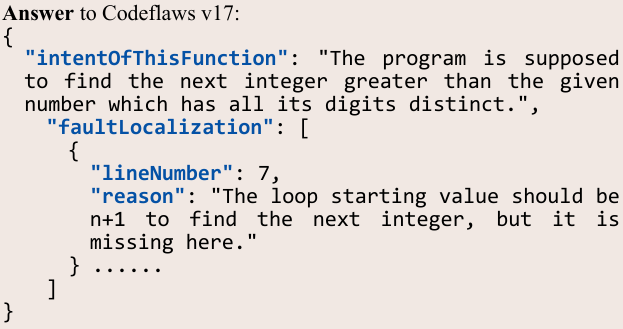}
\end{myleftline}
        
In this program, LLM-based FL outperforms traditional FL due to its advanced logical reasoning capabilities, enabling it to identify faults within the for loop. Traditional methods, constrained by their reliance on coverage-based suspiciousness calculations, lack the contextual insight needed to localize such logical faults.




\textbf{2) Disadvantages of LLM}

Despite the numerous advantages of LLMs in fault localization, they also exhibit certain limitations compared to traditional methods. For instance, in identifying specific types of faults, such as simple logical mistakes, LLMs may not match the accuracy of traditional approaches, which benefit from targeted algorithms. Additionally, the high computational cost of LLMs represents a significant drawback, as they require substantially more resources and time than traditional methods, potentially becoming a bottleneck in resource-constrained environments. Therefore, in practical applications, it is essential to balance the advantages of LLMs with their associated costs to select the most appropriate fault localization strategy. In the following, we will delve into a detailed analysis of the costs associated with LLMs to provide an evaluation of their practical utility in novice program fault localization.

\begin{table}[!htbp]
  \centering
    \caption{Average Costs per Program for Different LLMs}
    \begin{tabular}{lcc}
    \toprule
    \multicolumn{1}{c}{\textbf{Technique}} & \textbf{Monetary Cost (\$)} & \textbf{Time Cost (s)} \\
    \midrule
    \textbf{OpenAI o3} & 0.0152 & 52.4 \\
    \textbf{o1-preview} & 0.4859 & 96.4 \\
    \textbf{o1-mini} & 0.1497 & 43.2 \\
    \textbf{GPT-4o} & 0.0092 & 15.3 \\
    \textbf{GPT-4} & 0.0673 & 21.5 \\
    \textbf{GPT-3.5-Turbo} & 0.0034 & 7.3 \\
    \midrule
    \textbf{SBFL} & -     & 0.8323 \\
    \textbf{MBFL} & -     & 38.2789 \\
    \midrule 
    \end{tabular}%
  \label{tab:cost}
\end{table}%

\textbf{Cost Analysis:}
The computational cost of LLM-based fault localization is one of their significant drawbacks. Table~\ref{tab:cost} presents the average monetary and time costs per program for different LLMs, where ``Monetary Cost" is measured in dollars and ``Time Cost" in seconds. Compared to traditional static or dynamic analysis tools, the reasoning process of LLMs requires high-performance computing resources, particularly when dealing with large code fragments or complex faults. This high computational cost is primarily attributed to the large parameter size and high complexity of reasoning, especially when multi-turn interactions or CoT reasoning are employed to analyze issues in depth, leading to increased time and energy consumption. Moreover, for individual developers or small teams, relying on cloud-based services for LLMs incurs additional usage costs, limiting their applicability in resource-constrained scenarios. Although LLMs provide highly intelligent fault localization capabilities, their computational cost disadvantages may hinder their widespread adoption in low-resource environments.

\definecolor{customcolor}{HTML}{c8d7f0}

\begin{table*}[!htbp]
\centering
\caption{Common Fault Patterns and Their Explanations in Novice Programming}
\renewcommand{\arraystretch}{1.2}

\begin{tabular}{@{}p{0.48\textwidth}p{0.48\textwidth}@{}}
\noalign{\vskip 3pt} 
\centering\arraybackslash\textbf{Fault Code} & \centering\arraybackslash\textbf{Fault Explanation} \\
\noalign{\vskip 8pt} 
\begin{tcolorbox}[
  arc=4mm,
  boxrule=0pt,
  colback=customcolor!10,
  width=\textwidth,
  halign=center,
  valign=center,
  before skip=0pt,
  after skip=0pt,
  left=5pt,    
  right=2pt,   
  top=-4pt,    
  bottom=-4pt, 
]
\begin{minipage}[c]{0.47\textwidth}
\begin{lstlisting}[numbers=none]
74 if(state==1&&bit!=0)
75   func(s);
\end{lstlisting}
\end{minipage}%
\hfill%
\begin{minipage}[c]{0.47\textwidth}
\begin{tcolorbox}[
  arc=4mm,
  boxrule=0pt,
  colback=customcolor,frame empty,
  width=\linewidth,
  halign=left,
  valign=center,
  top=2pt,
  bottom=2pt
]
Potential logical fault in the condition check for calling the `func' function.
\end{tcolorbox}
\end{minipage}
\end{tcolorbox}\\[-7pt]

\begin{tcolorbox}[
  arc=4mm,
  boxrule=0pt,
  colback=white,
  width=\textwidth,
  halign=center,
  valign=center,
  before skip=0pt,
  after skip=0pt,
  left=5pt,    
  right=2pt,   
  top=-6pt,    
  bottom=-6pt, 
]
\begin{minipage}[c]{0.47\textwidth}
\begin{lstlisting}[numbers=none]
26 scanf("%d",&n);
27 if(n==0)
28   fun(0);
\end{lstlisting}
\end{minipage}%
\hfill%
\begin{minipage}[c]{0.47\textwidth}
\begin{tcolorbox}[
  arc=4mm,
  boxrule=0pt,
  colback=customcolor,frame empty,
  width=\linewidth,
  halign=left,
  valign=center,
  top=2pt,
  bottom=2pt
]
This line checks if `n' is 0 but doesn't handle it in the loop, which might lead to incorrect behavior.
\end{tcolorbox}
\end{minipage}
\end{tcolorbox}\\[-7pt]

\begin{tcolorbox}[
  arc=4mm,
  boxrule=0pt,
  colback=customcolor!10,
  width=\textwidth,
  halign=center,
  valign=center,
  before skip=0pt,
  after skip=0pt,
  left=5pt,    
  right=2pt,   
  top=-4pt,    
  bottom=-4pt, 
]
\begin{minipage}[c]{0.47\textwidth}
\begin{lstlisting}[numbers=none]
32 if(tut<x)
33   tut=x;
\end{lstlisting}
\end{minipage}%
\hfill%
\begin{minipage}[c]{0.47\textwidth}
\begin{tcolorbox}[
  arc=4mm,
  boxrule=0pt,
  colback=customcolor,frame empty,
  width=\linewidth,
  halign=left,
  valign=center,
  top=2pt,
  bottom=2pt
]
Potential fault due to inconsistency in handling even and odd positions.
\end{tcolorbox}
\end{minipage}
\end{tcolorbox}\\[-7pt]

\begin{tcolorbox}[
  arc=4mm,
  boxrule=0pt,
  colback=white,
  width=\textwidth,
  halign=center,
  valign=center,
  before skip=0pt,
  after skip=0pt,
  left=5pt,    
  right=2pt,   
  top=-6pt,    
  bottom=-6pt, 
]
\begin{minipage}[c]{0.47\textwidth}
\begin{lstlisting}[numbers=none, breaklines=true, basicstyle=\small\ttfamily]
27 ase 5 : {printf(""); break;}
28 case 7 : {printf("24 + 0 = 24"); break;}
29 default : {make24(n); break;}
\end{lstlisting}
\end{minipage}%
\hfill%
\begin{minipage}[c]{0.47\textwidth}
\begin{tcolorbox}[
  arc=3mm,
  boxrule=0pt,
  colback=customcolor,frame empty,
  width=\linewidth,
  halign=left,
  valign=center,
  top=2pt,
  bottom=2pt
]
Switch case handling might contain faults.
\end{tcolorbox}
\end{minipage}
\end{tcolorbox}\\[-7pt]

\begin{tcolorbox}[
  arc=4mm,
  boxrule=0pt,
  colback=customcolor!10,
  width=\textwidth,
  halign=center,
  valign=center,
  before skip=0pt,
  after skip=0pt,
  left=5pt,    
  right=2pt,   
  top=-6pt,    
  bottom=-6pt, 
]
\begin{minipage}[c]{0.47\textwidth}
\begin{lstlisting}[numbers=none, breaklines=true, basicstyle=\small\ttfamily]
4 int i,j,a[26]={0},b[26]={0},sum=0,flag=0;
5 char s1[1010],s2[1010];
6 scanf("%s",s1);
\end{lstlisting}
\end{minipage}%
\hfill%
\begin{minipage}[c]{0.47\textwidth}
\begin{tcolorbox}[
  arc=4mm,
  boxrule=0pt,
  colback=customcolor,frame empty,
  width=\linewidth,
  halign=left,
  valign=center,
  top=2pt,
  bottom=2pt
]
Array initialization with a fixed size may cause issues if input strings exceed the array size.
\end{tcolorbox}
\end{minipage}
\end{tcolorbox}\\[-7pt]

\begin{tcolorbox}[
  arc=4mm,
  boxrule=0pt,
  colback=white,
  width=\textwidth,
  halign=center,
  valign=center,
  before skip=0pt,
  after skip=0pt,
  left=5pt,    
  right=2pt,   
  top=-5pt,    
  bottom=-5pt, 
]
\begin{minipage}[c]{0.47\textwidth}
\begin{lstlisting}[numbers=none]
9 printf("-1");
10 return 0;
\end{lstlisting}
\end{minipage}%
\hfill%
\begin{minipage}[c]{0.47\textwidth}
\begin{tcolorbox}[
  arc=4mm,
  boxrule=0pt,
  colback=customcolor,frame empty,
  width=\linewidth,
  halign=left,
  valign=center,
  top=2pt,
  bottom=2pt
]
Printing -1 without linebreaks might affect the output.
\end{tcolorbox}
\end{minipage}
\end{tcolorbox}\\[-7pt]

\begin{tcolorbox}[
  arc=4mm,
  boxrule=0pt,
  colback=customcolor!10,
  width=\textwidth,
  halign=center,
  valign=center,
  before skip=0pt,
  after skip=0pt,
  left=5pt,    
  right=2pt,   
  top=-2pt,    
  bottom=-2pt, 
]
\begin{minipage}[c]{0.47\textwidth}
\begin{lstlisting}[numbers=none]
11 if(n%2==0||n==1)
\end{lstlisting}
\end{minipage}%
\hfill%
\begin{minipage}[c]{0.47\textwidth}
\begin{tcolorbox}[
  arc=4mm,
  boxrule=0pt,
  colback=customcolor,frame empty,
  width=\linewidth,
  halign=left,
  valign=center,
  top=2pt,
  bottom=2pt
]
Potential fault in the conditional logic for n\%2==0 , n==1, which may not cover all edge cases.
\end{tcolorbox}
\end{minipage}
\end{tcolorbox}\\[-7pt]

\begin{tcolorbox}[
  arc=4mm,
  boxrule=0pt,
  colback=white,
  width=\textwidth,
  halign=center,
  valign=center,
  before skip=0pt,
  after skip=0pt,
  left=5pt,    
  right=2pt,   
  top=-2pt,    
  bottom=-2pt, 
]
\begin{minipage}[c]{0.47\textwidth}
\begin{lstlisting}[numbers=none]
23 if(oc<2)
\end{lstlisting}
\end{minipage}%
\hfill%
\begin{minipage}[c]{0.47\textwidth}
\begin{tcolorbox}[
  arc=4mm,
  boxrule=0pt,
  colback=customcolor,frame empty,
  width=\linewidth,
  halign=left,
  valign=center,
  top=2pt,
  bottom=2pt
]
Potential fault: Insufficient check for the presence of only one odd character count. The condition may not cover all edge cases.
\end{tcolorbox}
\end{minipage}
\end{tcolorbox}\\[-7pt]

\begin{tcolorbox}[
  arc=4mm,
  boxrule=0pt,
  colback=customcolor!10,
  width=\textwidth,
  halign=center,
  valign=center,
  before skip=0pt,
  after skip=0pt,
  left=5pt,    
  right=2pt,   
  top=-6pt,    
  bottom=-6pt, 
]
\begin{minipage}[c]{0.47\textwidth}
\begin{lstlisting}[numbers=none]
27 if (e == 0)
28   printf("YES\n");
\end{lstlisting}
\end{minipage}%
\hfill%
\begin{minipage}[c]{0.47\textwidth}
\begin{tcolorbox}[
  arc=4mm,
  boxrule=0pt,
  colback=customcolor,frame empty,
  width=\linewidth,
  halign=left,
  valign=center,
  top=2pt,
  bottom=2pt
]
Final validation of input sequence may not be accurate.
\end{tcolorbox}
\end{minipage}
\end{tcolorbox}\\[-7pt]

\begin{tcolorbox}[
  arc=4mm,
  boxrule=0pt,
  colback=white,
  width=\textwidth,
  halign=center,
  valign=center,
  before skip=0pt,
  after skip=0pt,
  left=5pt,    
  right=2pt,   
  top=-6pt,    
  bottom=-6pt, 
]
\begin{minipage}[c]{0.47\textwidth}
\begin{lstlisting}[numbers=none]
55 S[j] = '\0';
56 printf("%s\n", S);
\end{lstlisting}
\end{minipage}%
\hfill%
\begin{minipage}[c]{0.47\textwidth}
\begin{tcolorbox}[
  arc=4mm,
  boxrule=0pt,
  colback=customcolor,frame empty,
  width=\linewidth,
  halign=left,
  valign=center,
  top=2pt,
  bottom=2pt
]
Modifying the source string `S' directly might lead to incorrect output in some cases.
\end{tcolorbox}
\end{minipage}
\end{tcolorbox}\\[-7pt]

\begin{tcolorbox}[
  arc=4mm,
  boxrule=0pt,
  colback=customcolor!10,
  width=\textwidth,
  halign=center,
  valign=center,
  before skip=0pt,
  after skip=0pt,
  left=5pt,    
  right=2pt,   
  top=-4pt,    
  bottom=-6pt, 
]
\begin{minipage}[c]{0.47\textwidth}
\begin{lstlisting}[numbers=none]
21 if(arr[i]>=cum){
22  happy++;
\end{lstlisting}
\end{minipage}%
\hfill%
\begin{minipage}[c]{0.47\textwidth}
\begin{tcolorbox}[
  arc=4mm,
  boxrule=0pt,
  colback=customcolor,frame empty,
  width=\linewidth,
  halign=left,
  valign=center,
  top=2pt,
  bottom=2pt
]
Potential logic fault, should compare arr[i] with cum + arr[i] for correct calculation
\end{tcolorbox}
\end{minipage}
\end{tcolorbox}\\[-7pt]

\begin{tcolorbox}[
  arc=4mm,
  boxrule=0pt,
  colback=white,
  width=\textwidth,
  halign=center,
  valign=center,
  before skip=0pt,
  after skip=0pt,
  left=5pt,    
  right=2pt,   
  top=-5pt,    
  bottom=-5pt, 
]
\begin{minipage}[c]{0.47\textwidth}
\begin{lstlisting}[numbers=none]
26 else if(a1[i] < a2[i]){
27  printf("second");
\end{lstlisting}
\end{minipage}%
\hfill%
\begin{minipage}[c]{0.47\textwidth}
\begin{tcolorbox}[
  arc=4mm,
  boxrule=0pt,
  colback=customcolor,frame empty,
  width=\linewidth,
  halign=left,
  valign=center,
  top=2pt,
  bottom=2pt
]
The loop may break prematurely due to the `break' statement, causing incorrect comparisons.
\end{tcolorbox}
\end{minipage}
\end{tcolorbox}\\[-7pt]

\begin{tcolorbox}[
  arc=4mm,
  boxrule=0pt,
  colback=customcolor!10,
  width=\textwidth,
  halign=center,
  valign=center,
  before skip=0pt,
  after skip=0pt,
  left=5pt,    
  right=2pt,   
  top=-6pt,    
  bottom=-6pt, 
]
\begin{minipage}[c]{0.47\textwidth}
\begin{lstlisting}[numbers=none]
13 else if(n[i]=='4')d*=3;
14 else if(n[i]=='5')d*=4;
15 else if(n[i]=='6')d*=2;
\end{lstlisting}
\end{minipage}%
\hfill%
\begin{minipage}[c]{0.47\textwidth}
\begin{tcolorbox}[
  arc=4mm,
  boxrule=0pt,
  colback=customcolor,frame empty,
  width=\linewidth,
  halign=left,
  valign=center,
  top=2pt,
  bottom=2pt
]
Mapping for `4' to multiplier 3 may be unintended; check if correct mapping intended.
\end{tcolorbox}
\end{minipage}
\end{tcolorbox}\\[-7pt]

\begin{tcolorbox}[
  arc=4mm,
  boxrule=0pt,
  colback=white,
  width=\textwidth,
  halign=center,
  valign=center,
  before skip=0pt,
  after skip=0pt,
  left=5pt,    
  right=2pt,   
  top=-6pt,    
  bottom=-6pt, 
]
\begin{minipage}[c]{0.47\textwidth}
\begin{lstlisting}[numbers=none]
25  count++;
26 else
27  a[s[i]-65]--;
\end{lstlisting}
\end{minipage}%
\hfill%
\begin{minipage}[c]{0.47\textwidth}
\begin{tcolorbox}[
  arc=4mm,
  boxrule=0pt,
  colback=customcolor,frame empty,
  width=\linewidth,
  halign=left,
  valign=center,
  top=2pt,
  bottom=2pt
]
Incrementing count without proper bounds checking or validation.
\end{tcolorbox}
\end{minipage}
\end{tcolorbox}\\[-7pt]

\end{tabular}
\label{tab:Common_Fault_Codes_and_Explanations}
\end{table*}

\subsection{\revise{Model Capability Paradox and Dataset Characteristics Analysis in Simple Fault Localization Tasks}}
\label{subsec:discuss-paradox}

\revise{From the results of RQ1, we observe that on the Codeflaws dataset, the Top-1 accuracy of OpenAI o3 is lower than GPT-4o. This result is counterintuitive, as the official product documentation suggests that iterative improvements in the model should lead to enhanced capabilities. This phenomenon is therefore inconsistent with the expected performance improvements. Although we recognize that such improvements do not always manifest uniformly across all tasks, we still expect more stable performance gains in common programming tasks. To further investigate this phenomenon, we select OpenAI o3 and the GPT-3.5-Turbo models from the GPT series covered in this paper, and evaluate them across multiple datasets using a consistent Top-N metric. In order to make the conclusion that the phenomenon on the Codeflaws dataset is special more convincing, we employ a one-sided Wilcoxon signed-rank test. The use of the Wilcoxon signed-rank test to test the effects of different technologies is commonly used in previous studies~\cite{wu2020fatoc}.}

\revise{The null hypothesis is formulated as:}

\revise{\textit{H}0. OpenAI o3 has a stronger fault localization effect in terms of Top-N metric than GPT-3.5-Turbo.}

\revise{We set the significance level at $\alpha$=0.05. A p-value lower than this threshold allows us to reject the null hypothesis. The results are shown in Table~\ref{tab:Wilcoxon Signed-rank Test Results across Different Datasets}.}

\begin{table}[!htbp]
  \caption{\revise{Wilcoxon Signed-rank Test Results across Different Datasets}}
    \resizebox{0.49\textwidth}{!}{
    \begin{tabular}{lcccccc}
    \toprule
    \multicolumn{1}{c}{\textbf{Dataset}} & \multicolumn{2}{c}{\textbf{Codeflaws}} & \multicolumn{2}{c}{\textbf{Condefects}} & \multicolumn{2}{c}{\textbf{BugT}} \\
    \midrule
    \textbf{Technique} & \textbf{o3} & \textbf{GPT-3.5} & \textbf{o3} & \textbf{GPT-3.5} & \textbf{o3} & \textbf{GPT-3.5} \\
    \midrule
    \textbf{Top-1} & 99    & 100   & 312   & 148   & 289   & 133 \\
    \midrule
    \textbf{Top-2} & 153   & 157   & 357   & 212   & 374   & 185 \\
    \midrule
    \textbf{Top-3} & 215   & 194   & 373   & 248   & 413   & 223 \\
    \midrule
    \textbf{Top-4} & 257   & 221   & 394   & 274   & 437   & 246 \\
    \midrule
    \textbf{Top-5} & 291   & 240   & 409   & 290   & 449   & 278 \\
    \midrule
    \textbf{p-value} & \multicolumn{2}{c}{0.15625} & \multicolumn{2}{c}{0.03125} & \multicolumn{2}{c}{0.03125} \\
    \bottomrule
    \end{tabular}%
    }
  \label{tab:Wilcoxon Signed-rank Test Results across Different Datasets}%
\end{table}%

\revise{From the table, we can see that on the Condefects and BugT datasets, OpenAI o3 has significantly stronger fault localization effect than GPT-3.5-Turbo at the 95\% confidence level. However, on the Codeflaws dataset, it cannot be considered that OpenAI o3 is better than GPT-3.5-Turbo, which makes our conjecture more convincing.}


To understand the factors influencing this unexpected trend, we manually selected and analyzed samples from the Codeflaws dataset where GPT-3.5-Turbo is able to localize the faults, but OpenAI o3 is not able to do so. We have listed the faulty code snippets and their explanations from all samples in Table ~\ref{tab:Common_Fault_Codes_and_Explanations}. We manually analyzed the fault explanations for all the samples. Characteristics of these faults reveal that common fault types include logical faults, algorithmic issues, and array/resource access problems. Most of the faults are related to incomplete assumptions or improper condition checks in the program design.


\revise{Although both GPT-3.5-Turbo and OpenAI o3 are black-box models and the exact reasons for their performance differences remain unclear, existing evidence suggests that the Codeflaws dataset may suffer from data leakage, unlike the other two datasets. During model iteration, OpenAI o3 is expected to have stronger generalization capabilities than GPT-3.5-Turbo. However, if Codeflaws contains problems that overlap with the training data, it may favor earlier models like GPT-3.5-Turbo, thereby leading to the counterintuitive observation that OpenAI o3 performs worse on this dataset.} Meanwhile,the architecture of the OpenAI o3 places a stronger emphasis on contextual reasoning and situational understanding\revise{~\cite{openai2025o3}}. When faced with faults, the OpenAI o3 considers more contextual information and the surrounding context during reasoning, attempting to understand the potential causes of the fault, especially when dealing with complex text. While this reasoning mechanism helps OpenAI o3 solve more complex tasks, excessive reasoning in the case of simple faults may lead to misdirection~\cite{yan2025recitationreasoningcuttingedgelanguage}, causing it to miss more direct fault localization.

In contrast, GPT-3.5-Turbo's response is more straightforward. It focuses more on obvious patterns and faults, especially simple boundary condition issues. For such problems, GPT-3.5-Turbo may not require much contextual reasoning to localize potential faults (such as handling n == 0). Its simplified reasoning mechanism makes GPT-3.5-Turbo more adept at handling these direct faults.

\subsection{\revise{Key Limitations from Our Study and Recommended Future Directions}}
\label{subsec:limitations-future}

\revise{Our study reveals key limitations in both LLM-based and traditional fault localization approaches when applied to novice programs.}

\revise{\textbf{LLM-based Method Limitations.} Our experiments identify two main challenges with current LLM approaches. First, high costs make advanced models like OpenAI o3 less practical for educational use compared to traditional methods. Second, our analysis reveals that advanced models can overthink simple faults, missing straightforward bugs that simpler approaches catch easily.}

\revise{\textbf{Traditional Method Limitations.} Our study also highlights key weaknesses in traditional approaches for novice programs. First, these methods suffer from insufficient accuracy in fault localization. Second, traditional methods cannot provide meaningful fault explanations to help novice programmers understand and learn from their mistakes.}

\revise{\textbf{Our Recommended Future Directions.} Based on our findings, we suggest several research paths. We recommend developing cost-effective models specifically designed for educational fault localization and smart reasoning systems that adjust their analysis based on fault complexity. We also suggest creating better explanation tools that help students learn from their mistakes.}

\revise{We particularly recommend hybrid approaches that combine the speed of traditional methods with the understanding capabilities of LLMs. Such systems could use traditional methods for initial screening while applying LLMs for deeper analysis and student-friendly explanations.}

\section{Threats to Validity}
\label{sec:Threats to Validity}


\textbf{Internal Validity.} The first threat to internal validity stems from possible faults in the program code. To address this, we conduct a thorough review of the code and perform small-scale tests. Furthermore, we retain essential data produced during code execution, which enables manual calculation and verification of the results, ensuring the program functions as intended.
The second internal validity is prompt design. Different prompts may elicit distinct responses from an LLM when addressing the same issue. To mitigate this, we conduct a series of ablation experiments to confirm that our carefully crafted prompts guide the LLMs to generate optimal results.
The third internal validity stems from the inherent randomness of LLMs. The same prompt may yield different responses across multiple runs. To mitigate this issue, we conduct repeated experiments and compute the average results, reducing the impact of randomness on our findings.

\textbf{External Validity.} The first external validity pertains to our dataset. The dataset, selected from Codeflaws, Condefects, and BugT consists of a carefully curated collection of real-world programming faults from Codeforces, Atcoder, and BuctOJ. This collection has been rigorously filtered to ensure a diverse range of fault types. \revise{We focus more on computer science students than on amateur self-learners, whose novice programs are usually written in C, C++ or Java~\cite{MANORAT2025100403}.}
For ease of analysis, we utilize these three language programs for our experiments.
The second external validity is dataset leakage. To mitigate the threat of dataset leakage, the Condefects and BugT dataset are designed for this purpose. It consists of the most recent faulty programs to minimize overlap with the training data used by LLMs.
Consequently, the potential threat posed by dataset leakage is limited.

\textbf{Construct Validity.}The first construct validity refers to the effectiveness of the evaluation metrics used in our study. We employ the Top-N metric to assess novice program fault localization performance, as it is widely adopted in our field. Given its extensive use in previous fault localization studies, this threat is limited. The second construct validity addresses the representativeness of the LLMs used for evaluating novice program fault localization.
By incorporating widely recognized models such as o1-preview and GPT-4, as well as open-source models like ChatGLM4, Llama3, and DeepSeekV3, we ensure a diverse range of model types.
This approach reduces the potential influence of specific model biases and mitigates this threat.

\section{Related Work}
\label{sec:Related Work}

\subsection{LLM-based Fault Localization}
Currently, research on leveraging LLMs for fault localization remains limited. Kang et al.~\cite{kang2023preliminary} introduced AutoFL, a fault localization approach utilizing GPT. AutoFL stands out by combining the function call capabilities of OpenAI's LLMs with custom-designed functions, enabling the model to explore and extract pertinent information from extensive source code repositories. Their evaluation of the Defects4J dataset revealed AutoFL's ability to localize 149 faults across 353 cases, demonstrating the promise of LLM-based fault localization.
Wu et al.~\cite{wu2023large} examined the performance of GPT-3.5-Turbo and GPT-4 in the context of fault localization. They meticulously crafted prompts, which included the target function, foundational instructions, and response formatting for GPT, and conducted experiments on the Defects4J dataset. Their study found that within a function-level context, GPT-4 delivered optimal fault localization results. However, when the code context expanded to the class level, GPT-4's performance declined, falling below the accuracy of the state-of-the-art baseline SmartFL. While LLMs may not always outperform traditional methods, their potential remains significant\revise{~\cite{10891926}}.

In our empirical study, we observe that fault localization research has predominantly focused on industrial applications, with limited attention given to novice programming education. The nature of novice programs, characterized by simpler faults and fewer tokens, makes them particularly well-suited for LLM-based fault localization.








\subsection{LLM-based Programming Feedback}

With the advent of large language models (LLMs), the landscape of educational technology, particularly in programming education \revise{is experiencing an revolution~\cite{10938596}}. These models have not only enhanced the quality of feedback provided to learners but also diversified the methods through which this feedback is generated and validated~\cite{phung2023generativeaiprogrammingeducation}. Recent advancements have seen LLMs being employed to simulate both tutor and student roles, enhancing the interactivity and precision of programming feedback.

For instance, Lohr et al.~\cite{lohr2024yourenottype} explored how LLMs can be effectively utilized to generate specific types of feedback for introductory programming tasks. They highlighted the transformation in feedback mechanisms from traditional deterministic methods to more dynamic, AI-generated responses, providing richer and more personalized feedback. This study emphasized the potential of LLMs to cater to various feedback dimensions such as correctness, specificity, and adaptiveness to the learner’s needs.

In a more focused approach, Phung et al.~\cite{phung2024automatinghumantutorstyleprogramming} developed a novel technique utilizing GPT-4 for generating hints and GPT-3.5-Turbo for validating these hints. Their system, aimed at automating human tutor-style feedback, incorporates symbolic information about failing test cases and possible fixes to enhance the relevancy and accuracy of the feedback provided. This dual-model approach not only boosts the generative quality of the hints but also employs an automatic validation mechanism to ensure the utility and applicability of the feedback in real-world learning scenarios.

\section{Conclusion}
\label{sec:Conclusion}

In this paper, we present a comprehensive investigation into the application of LLMs for novice programming fault localization.
Our systematic evaluation across different prompting strategies and problem difficulty levels reveals that LLMs generally outperform traditional approaches in this domain.
Closed-source LLMs like o1-preview and GPT-4, along with specialized open-source models such as DeepSeekR1, demonstrate significant advantages in accurately identifying faults in novice programs while providing explanations and guidance.

Specifically, our comparative analysis yields several detailed findings about LLM capabilities.
We demonstrate that model scale does not necessarily guarantee superior performance, with certain closed-source models occasionally underperforming despite having more advanced architectures.
Through overlapping line analysis, we confirm that different methods successfully identify unique fault types that others miss.
\revise{Furthermore, our findings reveal that LLM's fault explanations provide substantial assistance for novice programmers, demonstrating the significant reference value of LLMs in supporting beginners' fault comprehension and practical debugging processes.}

Building upon these findings, future work could explore the integration of LLMs with traditional fault localization techniques to create hybrid systems that leverage the complementary strengths of both approaches.
This integration would address the limitations identified in our analysis while maximizing the unique capabilities of each method for more comprehensive fault localization solutions in novice programming education.

\section*{Acknowledgment}
The work described in this paper is supported by the National Natural Science Foundation of China under Grant No.$61902015$.

\section*{Declaration of Generative AI and AI-assisted technologies in the writing process}
During the preparation of this work the author(s) used the ChatGPT of OpenAI in order to improve language and readability, with caution. After using this tool, the authors reviewed and edited the content as needed and take full responsibility for the content of the publication.

\bibliographystyle{elsarticle-num} 

\bibliography{references}

\end{sloppypar}
\end{document}